\documentclass[amssymb,amsfonts,amsmath,superscriptaddress,preprint]{revtex4-1}

\usepackage{graphicx,xcolor}
\usepackage{amsmath, bbm}
\usepackage{mathtools}
\usepackage{natbib}
\usepackage{epstopdf}
\usepackage{color}

\begin{document}

\title{Turning a molecule into a coherent two-level quantum system\vspace{0cm}}
\author{Daqing Wang}%
\affiliation{Max Planck Institute for the Science of Light, 91058 Erlangen, Germany}
 \author{Hrishikesh Kelkar}%
\affiliation{Max Planck Institute for the Science of Light, 91058 Erlangen, Germany}
\author{Diego Martin-Cano}%
\affiliation{Max Planck Institute for the Science of Light, 91058 Erlangen, Germany}
\author{Dominik Rattenbacher}%
\affiliation{Max Planck Institute for the Science of Light, 91058 Erlangen, Germany}
\author{Alexey Shkarin}%
\affiliation{Max Planck Institute for the Science of Light, 91058 Erlangen, Germany}
\author{Tobias Utikal}%
\affiliation{Max Planck Institute for the Science of Light, 91058 Erlangen, Germany}
\author{Stephan G\"otzinger}%
\affiliation{Department of Physics, Friedrich Alexander University, 91058 Erlangen, Germany\vspace{2.5cm}}
\affiliation{Max Planck Institute for the Science of Light, 91058 Erlangen, Germany}
\author{Vahid Sandoghdar}
\affiliation{Max Planck Institute for the Science of Light, 91058 Erlangen, Germany}
\affiliation{Department of Physics, Friedrich Alexander University, 91058 Erlangen, Germany\vspace{2.5cm}}

\begin{abstract}
\vspace{0.5cm}
Molecules are ubiquitous in natural phenomena and man-made products, but their use in quantum optical applications has been hampered by incoherent internal vibrations and other phononic interactions with their environment. We have now succeeded in turning an organic molecule into a coherent two-level quantum system by placing it in an optical microcavity. This allows several unprecedented observations such as 99\% extinction of a laser beam by a single molecule, saturation with less than 0.5 photon, and nonclassical generation of few-photon super-bunched light. Furthermore, we demonstrate efficient interaction of the molecule-microcavity system with single photons generated by a second molecule in a distant laboratory. Our achievements pave the way for linear and nonlinear quantum photonic circuits based on organic platforms.  
\end{abstract} 

\maketitle

\section{Introduction}
Molecules provide very compact quantum systems that host well-defined transitions, ranging from the microwave to the ultraviolet domains associated with their rotational, vibrational and electronic states. In addition, these intrinsic mechanical and electronic degrees of freedom can be coupled through various well-defined transitions and selection rules. Indeed, molecular systems have attracted renewed attention within the community of quantum physics both in the gas\,\cite{Spaun:16, Truppe:17} and condensed\,\cite{Lidzey:00, Hakala:09, Schwartz:11, Chikkaraddy:16, Polisseni:16, Zhang:17, Skoff:18} phases. In the former case, molecules possess long-lived vibrational levels and well-resolved rotational transitions, but their cooling and trapping are difficult so that access to single molecules has only very recently being explored\,\cite{Liu:18}. On the other hand, while addressing single molecules in solids has been feasible with high spatial and spectral resolutions for nearly three decades, a substantial degree of decoherence remains in this system due to phononic couplings\,\cite{SMbook}.

The ground vibrational level of the electronic excited state ($|e, v=0\rangle$) in a dye molecule can couple to $|g, v=0\rangle$ and $|g, v\neq 0\rangle$ in the ground state following the Franck-Condon principle (see Fig.\,\ref{schematics}a). When embedded in a solid, each of these transitions entails a zero-phonon line (ZPL) and a phonon wing caused by coupling to matrix phonons (Debye-Waller factor). In the case of polycyclic aromatic hydrocarbons (PAH), the ZPL connecting $|g, v=0\rangle$ and $|e, v=0\rangle$ (00ZPL) can be narrowed by about $10^5$ folds to the Fourier limit when cooled to liquid helium temperatures. Nevertheless, the decay of $|e, v=0\rangle$ via $|g, v\ne 0\rangle$ levels and the subsequent fast relaxation of the latter states give rise to decoherence, making phase sensitive and nonlinear quantum operations inefficient\,\cite{Pototschnig:11,Maser:16}. One way to counter this decoherence is to enhance the 00ZPL in a selective manner and, thus, modify the branching ratio out of $|e, v=0\rangle$.  

In the past decade, there have been many efforts to enhance the radiative properties of molecules by plasmonic nanostructures\,\cite{Kuehn:06, Chikkaraddy:16}. However, the large bandwidth of plasmon resonances does not allow for selective addressing of narrow transitions. To remedy this, one can use optical microcavities to enhance molecular ZPLs by a substantial Purcell factor, $F=\frac{3}{4\pi^2}\frac{Q\lambda^3}{V}\gg1$ \,\cite{book-Berman94}. First attempts in this direction have indeed been reported\,\cite{Norris:97, Steiner:07, Chizhik:09, Wang:17}, but the results fall short of notable enhancements. A successful laboratory realization needs to consider and tackle several technical issues, especially in regard to the microcavity design. While monolithic microcavities are fairly robust, they are difficult to tune and not always compatible with the material of the quantum emitter. Open Fabry-Perot resonators, on the other hand, are difficult to stabilize in the cryostat but are conveniently adjustable and can be more easily combined with various materials. In this work, we employ an open, tunable and scannable Fabry-Perot microcavity with very small mode volume ($V$) and moderate quality factor ($Q$) \,\cite{Toninelli:10, Kelkar:15, Wang:17}.  

\section{Experimental}

\begin{figure}
\centering
\includegraphics[width=12 cm]{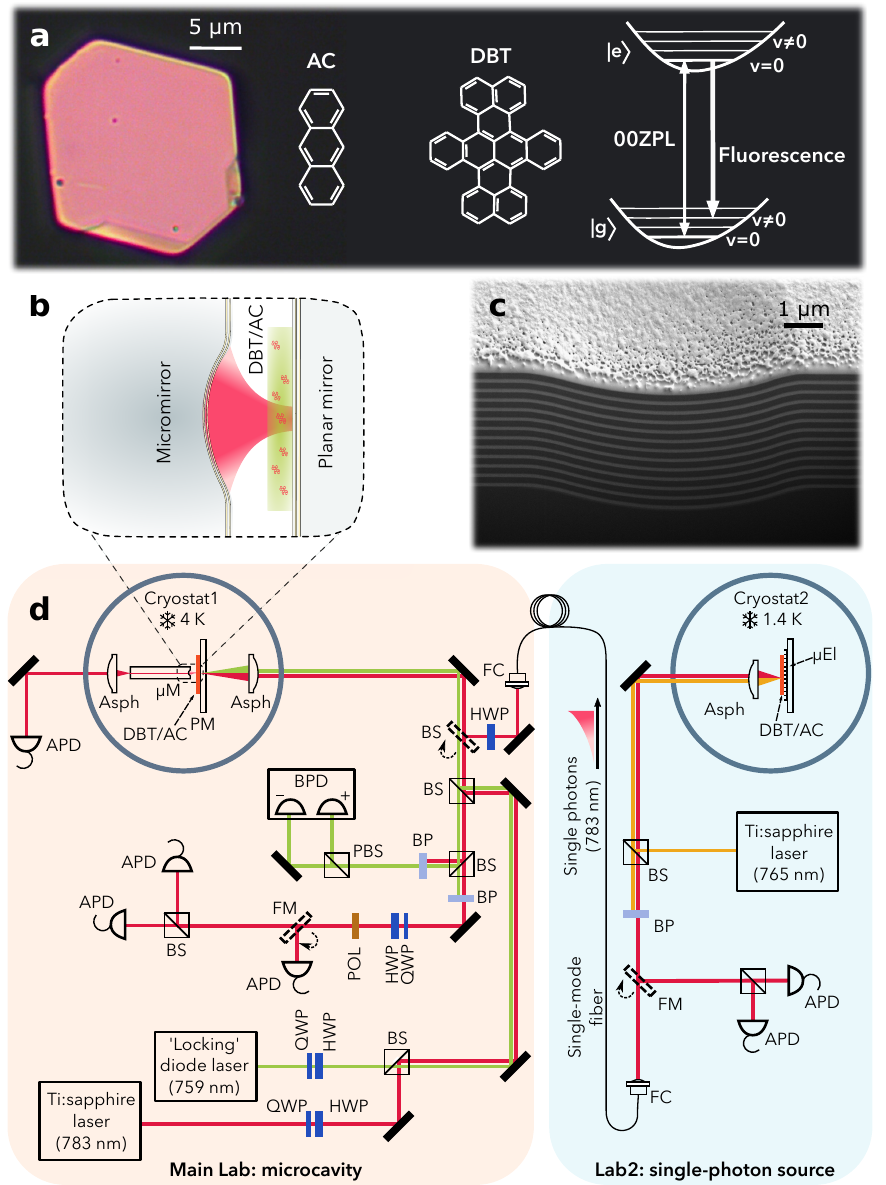}
\caption{\text{Schematics of the experimental arrangements.} \textbf{a,} An optical microscope image of a thin anthracene (AC) crystal, its molecular structures together with that of DBT, and the energy level scheme of the latter. \textbf{b,} Sketch of molecules embedded in a thin AC crystal placed in a microcavity. \textbf{c,} Electron microscope cross sectional image of the multilayer coating of the curved micromirror with radius of curvature of $10\,\mu$m after being cut by a focused ion beam. The apparent surface roughness is caused by the metallic coating necessary for electron microscopy and is absent on the mirror used in the measurements. \textbf{d,} Overview of the experimental setups in two different laboratories.} \label{schematics} 
\end{figure}

The method of choice for selecting single quantum emitters embedded in a solid is to exploit the inherent spectral inhomogeneity of such a system\,\cite{SMbook}. By operating at low enough temperatures, the homogenous linewidths of individual emitters become so narrow that they no longer overlap. Thus, a narrow-band laser beam can address the 00ZPLs of the various emitters located in the illumination volume one by one. In our experiment, we used dibenzoterrylene (DBT) of the PAH family embedded in a thin anthracene (AC) crystal (see Fig.\,\ref{schematics}a). The 00ZPL of DBT:AC lies in the interval $783-785$\,nm and can become as narrow as its Fourier limit of about 40\,MHz at $T \lesssim 4$\,K \,\cite{Nicolet:07}. 

To produce a Fabry-Perot cavity, we fabricated a curved micromirror at the end of an optical fiber and used a planar mirror \,\cite{Kelkar:15, Wang:17}, both coated with a dielectric multilayer (see Fig.\,\ref{schematics}b,c). The anthracene-filled microcavity exhibited an optical length of 4.7\,$\mu$m, cavity mode volume of $4.4\lambda^3$, finesse of 19,000, and $Q$-factor of 230,000. When placed in our helium exchange gas cryostat at 4\,K, residual vibrations broadened the line. This resulted in $Q=120,000$, deduced from the full width at half-maximum (FWHM) $\kappa/2\pi=3.3$\,GHz of a Voigt profile, which was composed of a Lorentzian resonance of FWHM=1.7\,GHz and a Gaussian broadening of FWHM=2.3\,GHz. In Fig.\,\ref{schematics}d, we display an overview of the experimental arrangement. 

\section{Results}

\subsection{Transmission and reflection of a laser beam}
\begin{figure}
\centering
\includegraphics[width=7.75 cm]{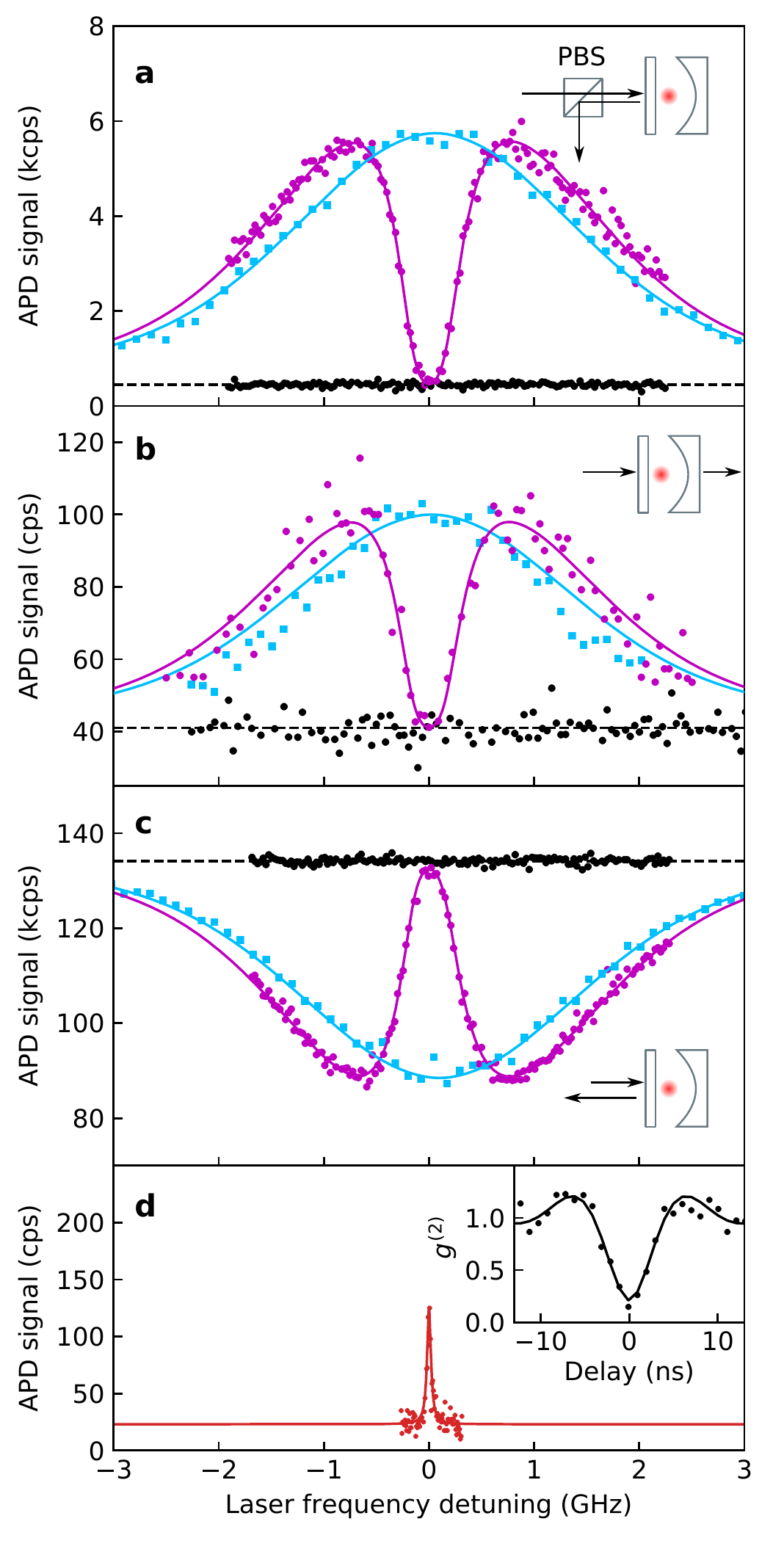}
\caption{\text{Resonant response of the molecule-microcavity composite to a laser beam.} \textbf{a,} Transmission spectra of the cavity with (magenta) and without (blue) the molecular contribution in units of counts per second (cps). The measurements were performed in cross-polarized mode. Black symbols show the intensity of the laser beam when the cavity resonance was detuned about 20\,GHz, acting as a reference level. A single molecule interrupts the cavity transmission by 99\%. \textbf{b,} Same as in (a) but for the direct transmission through the micromirror fabricated at the end of the optical fiber. \textbf{c,} Reflection spectra recorded on the same molecule as in (a,b). \textbf{d,} Fluorescence excitation spectrum of the same molecule recorded far detuned from the cavity resonance. The inset shows the second-order intensity autocorrelation function, verifying that this light is antibunched. See text for the explanation of the theoretical fits (solid curves).}
\label{resonant-spectra} 
\end{figure}

\hfill \break 
The presence of an emitter inside a cavity modifies the interference of the fields that result from the reflections between its mirrors. To probe the intracavity field, we exploit the cross-polarized reflection (CPR) that is generated by the birefringence of our cavity, providing an equivalent measure to a transmission recording\,\cite{Wang:17}. The blue symbols in Fig.\,\ref{resonant-spectra}a display a CPR spectrum in the absence of coupling to a molecule, and the blue solid curve shows a fit to these data using a Voigt profile. The black symbols in this figure represent a reference CPR signal when the cavity frequency was detuned by about 20\,GHz.

To examine the effect of a single molecule on the optical response of the microcavity, we tuned the resonance of the latter through the inhomogeneous band of DBT:AC and searched for the signature of molecular resonances directly in the cavity CPR spectrum while scanning the laser frequency. The magenta symbols in Fig.\,\ref{resonant-spectra}a present an example, where the cavity transmission drops by 99\% when it becomes resonant with a single molecule. In Fig.\,\ref{resonant-spectra}b, we also display a direct transmission measurement recorded through the micromirror (see Fig.\,\ref{schematics}d). All features in Figs.\,\ref{resonant-spectra}a and \ref{resonant-spectra}b agree, as confirmed by the high quality of the fits using common parameters. However, the signal-to-noise ratio (SNR) is lower in (b) because the numerical aperture of the single-mode fiber holding the micromirror does not match the cavity mode, thus, resulting in a weak signal. 

Next, we investigated the remarkably large effect of a single molecule on an incident laser beam further by measuring the direct reflection of the system. The blue symbols in Fig.\,\ref{resonant-spectra}c show the central part of the cavity resonance measured in the absence of molecular coupling. We note that the cavity resonance does not dip to zero due to imperfect mode matching and remaining vibrations. Nevertheless, the magenta symbols show that when the cavity is brought into resonance with a single molecule, the reflection dip vanishes. We shall present a quantitative discussion of the spectra in Fig. \ref{resonant-spectra}a-c as well as the solid theory curves used to fit them shortly. 

\subsection{Cavity QED modifications and spectral analysis}
\hfill \break 
The data in Fig.\,\ref{resonant-spectra}a-c let us deduce the FWHM of the molecular 00ZPL to be $604\pm21$\,MHz under coupling to the microcavity. To determine the linewidth of the very same molecule without the influence of the cavity, we detuned the resonance of the latter and recorded a red-shifted fluorescence signal from the molecule as a function of the excitation frequency. The outcome shown in Fig.\,\ref{resonant-spectra}d reveals FWHM=$44\pm5$\,MHz, which is in the range of the values for the bulk DBT:AC system\,\cite{Nicolet:07}. The weaker signal in this case stems from the difficulty of extracting the broad fluorescence through the higher-order transverse modes of the cavity. Photon antibunching of the fluorescence signal recorded via intensity autocorrelation ($g^{(2)}(\tau)$) confirms that it originates from only one molecule (see inset of Fig.\,\ref{resonant-spectra}d). 

The linewidth of an unperturbed molecule can be expressed as $\gamma^0=\gamma^0_\text{zpl}+\gamma_\text{red}$, where $\gamma^0_\text{zpl}$ stands for the decay rate of the excited state $|e, v=0\rangle$ into the 00ZPL channel, and $\gamma_\text{red}$ denotes the contributions of all red-shifted emission, including phonon wings and vibrational decay paths. When the cavity is resonant with the 00ZPL, the component of $\gamma^0_\text{zpl}$ emitted into the cavity mode is enhanced by the Purcell factor $F$, yielding $\gamma'_\text{zpl}=(1+F)\gamma^0_\text{zpl}$. Hence, considering that $\gamma_\text{red}\approx2\gamma^0_\text{zpl}$ for DBT:AC\,\,\cite{Trebbia:09}, we can write the modified decay rate of the excited state as $\gamma'=\gamma_\text{red}+\gamma'_\text{zpl}=(3+F)\gamma^0_\text{zpl}$. We, thus, deduce from our experimental findings of $\gamma^0/2\pi= 44\pm5$ MHz and $\gamma'/2\pi = 604\pm21$ MHz a Purcell factor of $F=38\pm 5$. 

A very useful measure for the efficiency of emitter-cavity coupling is the $\beta$-factor defined as the ratio of the power emitted into the cavity mode and the total emitted power\,\cite{Petermann:79}. The $\beta$-factor associated with an ideal two-level atom can be computed as $\frac{F}{F+1}$ and would correspond to $97.4\pm0.3\%$ for our cavity. To assess the overall degree of coherence for the resonant interaction between a DBT molecule and an incoming light field, however, we also have to account for losses to the red-shifted channels. Therefore, we arrive at $\beta=\frac{\gamma^0_\text{zpl}F}{\gamma'}=\frac{F}{F+3}=93\%$. 

The strong modification of the molecular emission on the 00ZPL changes its branching ratio $\alpha$, defined as the fraction of the power in the 00ZPL to the overall emission from the excited state. Our results demonstrate a modification from $\alpha\sim33\%$ for bulk DBT:AC to $\alpha'=\frac{\gamma'_\text{zpl}}{\gamma'}=\frac{F+1}{F+3}=95\%$ in the cavity. This implies that we have successfully converted a molecule to a two-level quantum system to within 95\%. The obtained high values of $\alpha$ and $\beta$ have immediate consequences for the efficiency of coherent linear and nonlinear processes at the single-molecule level\,\cite{Pototschnig:11,Maser:16}. 

\begin{figure}
\vspace{-1cm}
\centering
\includegraphics[width=11 cm]{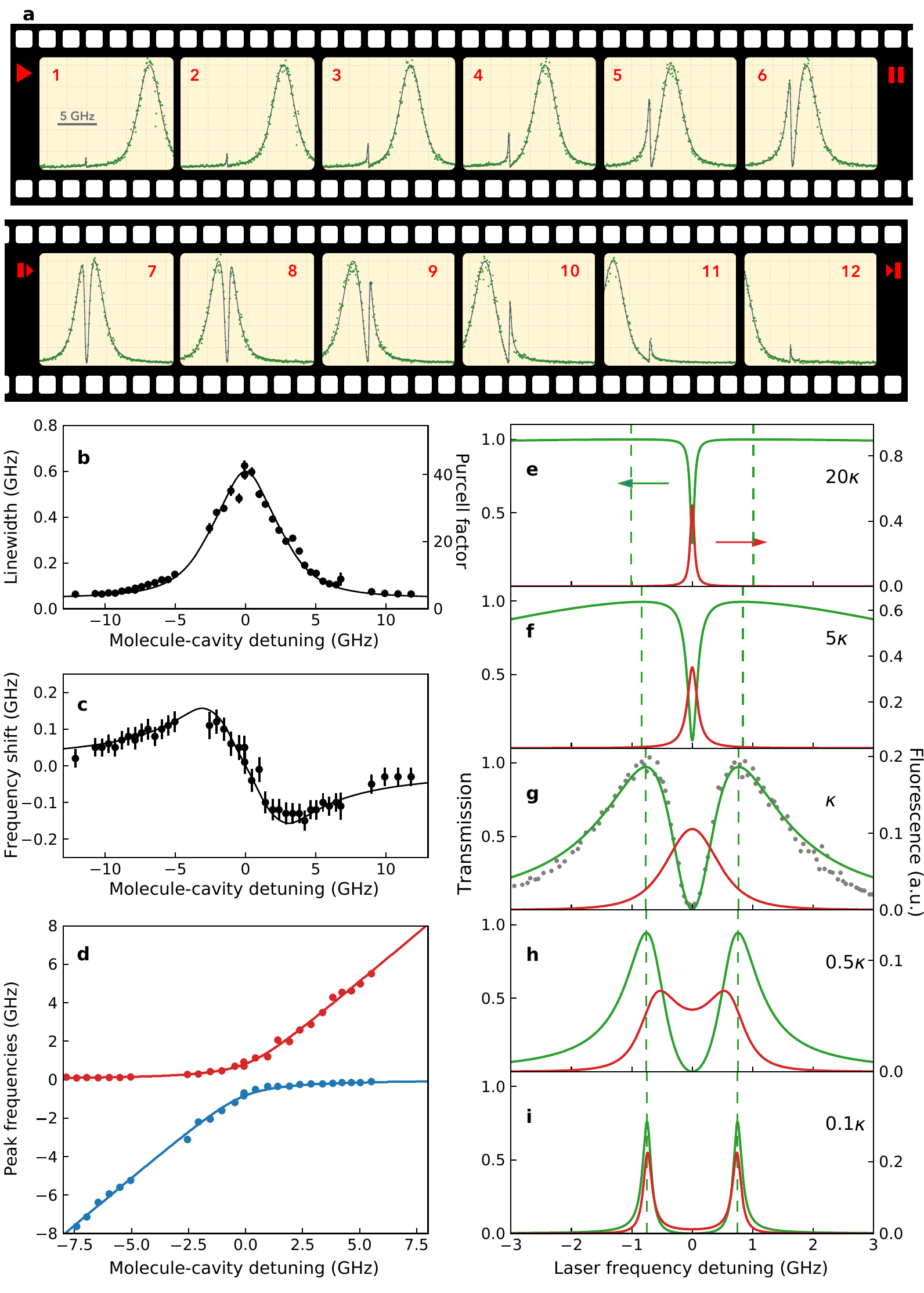}
\caption{\text{Frequency-detuned response of the molecule-cavity system.} \textbf{a,} Series of transmission spectra for different molecule-cavity frequency detunings. \textbf{b,} Linewidth of the molecular resonance and measured Purcell factor as a function of molecule-cavity frequency detuning. \textbf{c,} Frequency shift of the molecular resonance (the modified Lamb shift) as a function of molecule-cavity frequency detuning. \textbf{d,} Positions of the peaks in the transmission spectrum as a function of molecule-cavity frequency detuning. The solid curves show fits obtained from rigorous theoretical calculations. \textbf{e-i,} Calculated transmission (green) and fluorescence (red) spectra of a coupled system for decreasing degree of cavity loss. The legend in each figure denotes the cavity FWHM in terms of our experimental linewidth $\kappa$. The symbols in (g) represent the experimental CPR spectrum. The fit quality here is not as good because the contribution of vibrations to a Voigt profile are not taken into account in the calculated spectra. The dotted vertical lines displays the positions of the maxima in the transmission spectra.}
\label{detuned-spectra}
\end{figure}

To investigate the radiative modifications further, we took advantage of the axial tunability of our microcavity and recorded a series of CPR spectra at different molecule-cavity frequency detunings. Figure\,\ref{detuned-spectra}a demonstrates the evolution of the molecule-cavity spectral modifications, providing a wealth of quantitative data and a thorough comparison between experiment and theory. In a first simple approach, we fit the observed Fano-like dispersive line shapes using a generalized Lorentzian function. Figure\,\ref{detuned-spectra}b nicely traces the linewidth of the molecular resonance (left vertical axis) and the corresponding Purcell factor (right vertical axis) as a function of the cavity frequency detuning. The data in Fig.\,\ref{detuned-spectra}a can also be analyzed by considering a rigorous theory that treats the interaction of an incoming field with both the molecule and the cavity on the same footing\,\cite{Auffeves:07}. The solid curves in Figs.\,\ref{resonant-spectra}a-c and \ref{detuned-spectra}a show fits to the experimental data with excellent agreement using such a model. 

Next, we analyzed the influence of the cavity coupling on the center frequency of the molecular resonance. Figure\,\ref{detuned-spectra}c plots the latter as a function of the cavity-molecule frequency detuning and reveals frequency shifts by up to about $\pm150$\,MHz towards blue or red, depending on the sign of the detuning. This can be interpreted as a change in the contribution of vacuum fluctuations to the absolute value of the 00ZPL, i.e. of the Lamb shift\,\cite{book-Berman94, Heinzen:87}. Perhaps somewhat nonintuitively, the correction disappears when the cavity is resonant with the molecular line and peaks at the largest slope of the resonance profile. This behavior is similar to the well-known AC stark shift proportional to the laser-atom detuning\,\cite{cohen-book}.  

To place our parameters in the context of weak and strong coupling regimes, in Fig.\,\ref{detuned-spectra}e-i we present calculated transmission (green) and fluorescence (red) spectra for various cavity linewidths. While the fluorescence spectra clearly show the onset of a line splitting for higher finesse cavities, the splitting between the maxima in the transmission spectrum is nearly independent of the cavity finesse (see dotted vertical lines). The symbols in Fig.\,\ref{detuned-spectra}g show that the experimental data correspond to the transitional regime where the spectrum changes its character from a molecular extinction dip on a broad cavity resonance to one with two split polaritonic resonance profiles at par. The fluorescence spectrum in this region is shortly before bifurcation into two maxima, a signature of strong coupling\,\cite{book-Berman94}. A quantitative measure for the onset of strong coupling can be formulated by the exceptional point, where the cavity-molecule coupling rate $g$ satisfies $g_{\rm ep}=\lvert \kappa-\gamma \rvert/4$ \,\cite{Choi:10}. To determine $g$ for our experiment, in Fig.\,\ref{detuned-spectra}d we plot the frequencies of the two maxima that arise in the spectra of Fig.\,\ref{detuned-spectra}a. The splitting at zero detuning directly equals $2g$, yielding $g=0.79 \pm 0.3$\,GHz. Comparison of this value with $g_{\rm ep}=0.82$\,GHz computed for our system confirms that our experiment is situated right at the onset of strong coupling.  

A convenient parameter that connects $g$, $\kappa$ and $\gamma$ is the cooperativity factor $C=\frac{4g^2}{\kappa\gamma}$. For a two-level atom, $C$ and $F$ are equivalent, but one has to distinguish between them when dealing with emitters that support multichannel decay. To this end, our measured value of $F=38$ reports on the enhancement of $\gamma^0_{\rm zpl}$ as a well-defined dipolar transition into a single mode of a cavity. The expression of $C$, on the other hand, reports on the degree of coherence in the interaction of  the molecule as a whole with a photon in the cavity. Thus, to estimate $C$, we use the total decay rate $\gamma^0$ so as to account for the internal loss of coherence through the red-shifted emission paths, arriving at $C=12.7$. We note that, in fact, our microcavity values of $Q$ and $V$ let us expect a much higher Purcell factor $F\sim350$ and, thus, even stronger couplings than reported here. We attribute the discrepancy between the measured and predicted Purcell factors to the suboptimal position and orientation of the molecule with respect to the cavity electric field.

\subsection{Phase Shift}
\hfill \break 
The phase shift imprinted by a quantum emitter on a light beam can report on the state of the emitter in a nondestructive fashion. Previous experiments have demonstrated phase shifts of about three degrees applied to a focused laser beam by single molecules in a crystal\,\cite{Pototschnig:11}. Considering the high cooperativity and coupling efficiency of our system, we should now expect a much larger phase shift. To explore this, we examined the CPR of a laser beam from the cavity, following the protocol described in Ref.\,\,\cite{Pototschnig:11}. 

\begin{figure}
\centering
\includegraphics[width=10 cm]{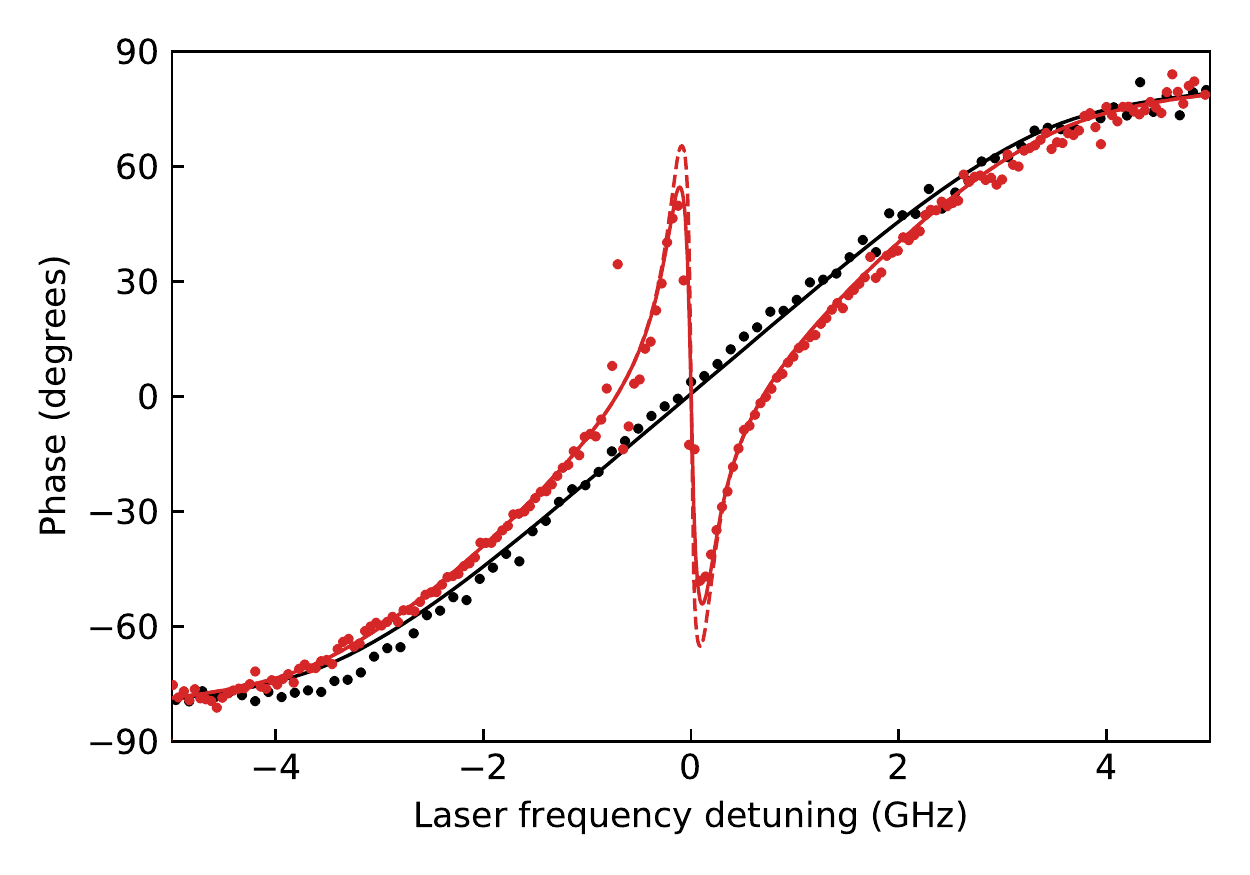}
\caption{\text{Large phase shift of a laser beam by a single molecule.} Measured phase shifts of a laser beam after interacting with a cavity without (black symbols) and with (red symbols) a single molecule. The curves show the theoretical fits.}
\label{phase}
\end{figure}

The black symbols in Fig.\,\ref{phase} show the phase shift affected by the microcavity alone as the laser frequency was scanned across its resonance in the absence of a molecule. The red symbols in that figure display the recorded phase shift of the laser beam under the influence of a single molecule. The solid and dashed curves signify theoretical fits with and without the consideration of power broadening, respectively, allowing us to deduce phase shifts up to $\phi=\pm 66^\circ$. 

\subsection{Photon Bunching and single-photon nonlinearity}
\begin{figure}
\centering
\includegraphics[width=12 cm]{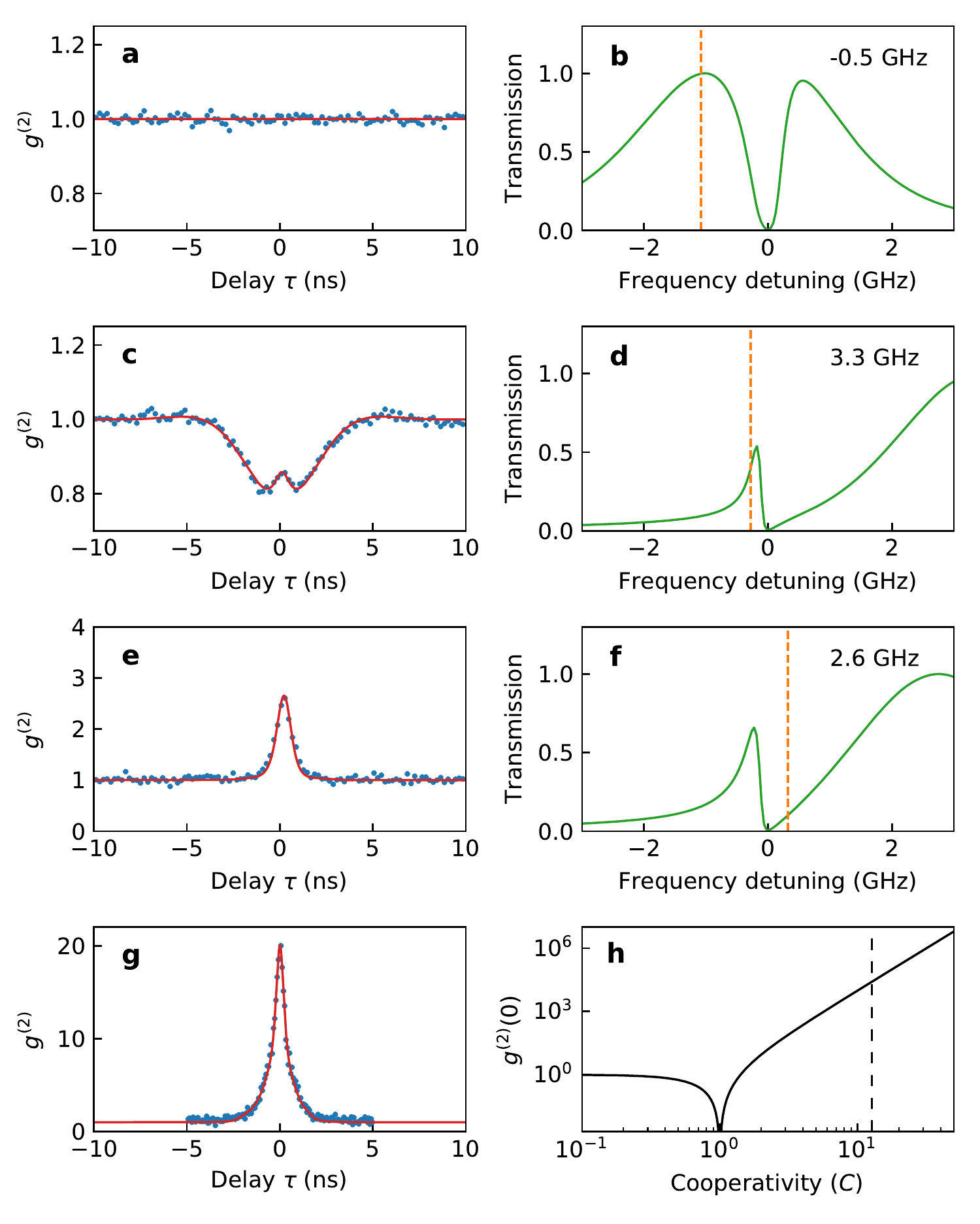}
\caption{\text{Strong modification of photon statistics.} \textbf{a, c, e,} Intensity autocorrelation $g^{(2)}{(\tau)}$ of the cross-polarized reflection (CPR) of a laser beam from the molecule-cavity composite for different molecule-cavity and laser frequencies. \textbf{b, d, f,} The CPR spectra corresponding to the measurements shown in (a), (c), (e). The orange dashed lines denote the laser frequency in each case. Molecule-cavity detuning is displayed as legend in each plot. \textbf{g,} Photon bunching corresponding to $g^{(2)}{(0)}=21$ at the molecular resonance in the situation of Fig.\,\ref{resonant-spectra}. The red curves in (a-g) show the theoretical fits. \textbf{h,} Theoretical predictions of $g^{(2)}{(0)}$ as a function of cooperativity $C$. The dashed line depicts the experimental parameter used in (g).}
\label{bunching}
\end{figure}

\hfill \break 
The photon statistics of a laser beam and of a quantum emitter take on very different forms, characterized by intensity autocorrelations $g^{(2)}(\tau)=1$ and $g^{(2)}(0)=0$, respectively. We now show that an efficient coupling between laser light and a molecule can result in highly nontrivial statistics of the emerging photons\,\cite{Rice:88}. Parts a,c,e, and g of Fig.\,\ref{bunching} display $g^{(2)}(\tau)$ measurements on a laser beam after interaction with the molecule-cavity system at different frequency detunings (see Figs.\,\ref{bunching}b,d,f). The outcome $g^{(2)}(\tau)=1$ in Fig.\,\ref{bunching}a reveals that the lower polariton branch in this case (see Fig.\,\ref{bunching}b) has a laser-like nature. The measurement shown in Fig.\,\ref{bunching}c, on the other hand, presents a nontrivial case of antibunching for the molecule-like branch of the spectrum in Fig.\,\ref{bunching}d. This antibunching results from the nonclassical interference of the molecular scattering with the intracavity field and provides evidence for the dipole quadrature  squeezing, which was recently detected in cavities\,\cite{Ourjoumtsev2011} and in free space\,\cite{Schulte2015}. 

Figure\,\ref{bunching}e shows that tuning the laser frequency by a bit more than a linewidth (see Fig.\,\ref{bunching}f) changes the behaviour completely to a \textit{bunching} effect. This phenomenon stems from the selective scattering of single-photon components from the Poisson distribution of photons in the incident laser beam\,\cite{Rice:88}, yielding a super-bunched few-photon state\,\cite{book-Ficek}. As displayed in Fig.\,\ref{bunching}g, this effect is maximized at the centre of the resonant molecule-cavity spectrum shown in Fig.\,\ref{resonant-spectra}a. Fitting the data by the theoretical model described in Ref.\,\,\cite{Carmichael:91} lets us deduce $g^{(2)}(0)=21$, which is among the largest photon bunchings reported to date for a single emitter\,\cite{Bennett:16,Snijders:16}. In fact, calculations in Fig.\,\ref{bunching}h show that for $C=12.7$ one expects $g^{(2)}(0)$ to reach as high as $2.5\times10^4$ if the molecule, cavity and laser frequencies coincide. The discrepancy with the measured value of 21 is due to the limited detector response time of 50\,ps and the residual background light.

The underlying mechanism of the phenomena observed above is that the molecule responds to only one photon at a time. This feature is also responsible for the intrinsic nonlinearity of an atom or molecule, which in turn leads to saturation as the excitation power is increased. We studied the nonlinear response of the cavity-coupled molecule by examining the extinction signal\,\cite{Wrigge:08}. We find that we reach the saturation parameter of $S=1$ for a very low power of 420\,pW coupled to the cavity, corresponding to only 0.44 photons per excited state lifetime of 264 ps. This result indicates that the operation regime of our experiment not only provides nearly perfect coupling in the weak excitation limit, but it also opens doors for efficient few-photon nonlinear operations\,\cite{Chang:14,Maser:16}. 

\subsection{Single-Photon Reflection}
\begin{figure}
\centering
\includegraphics[width=9.5 cm]{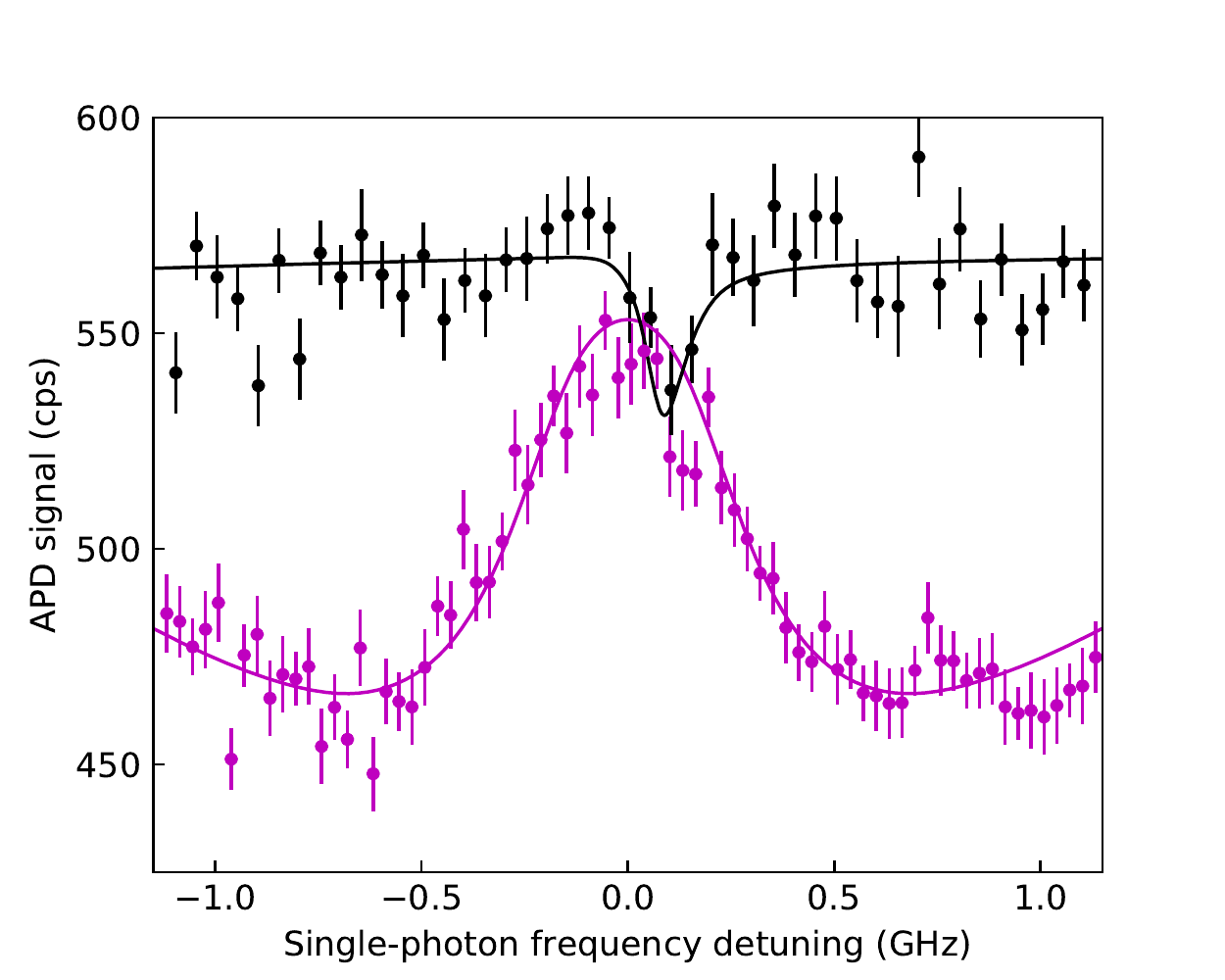}
\caption{\text{Reflection of single photons.} Reflection spectrum of the molecule-cavity system when single photons from another molecule in a different laboratory were impinged on it (magenta). Single photon budget: 6 kcps (kilo counts per second) out of the fiber in the microcavity lab, 3 kcps incident onto the cavity, 500 cps on the final detector. The black symbols show a spectrum recorded when the cavity resonance was detuned by 6\,GHz. Note that our freedom to change the frequency of the single photons via Stark effect on the ``source" molecule is limited compared to the case of a laser beam in Fig.\,\ref{resonant-spectra}. The small dip that is slightly shifted from the origin denotes the interference of the photons scattered by the target molecule with the component directly reflected at the flat mirror.}  
\label{single-photons}
\end{figure}

\hfill \break 
The ultimate frontier of light-matter interaction requires efficient coupling of a single photon and a single quantum emitter. To demonstrate such a ``gedanken" experiment, we used a second molecule located in a different laboratory (see Fig.\,\ref{schematics}d) as a source of narrow-band single photons (see Ref.\,\,\cite{Rezus:12} for details). The resulting stream of 30,000 photons per second with a FWHM linewidth of 41\,MHz was coupled to a single-mode fiber and sent to the laboratory housing the microcavity. The magenta symbols in Fig.\,\ref{single-photons} present the reflection spectrum of this single-photon stream when the cavity was tuned to resonance with the 00ZPL of the ``target" molecule. While the count rate and the shot-noise-limited SNR are lower, the signal reproduces our findings in Fig.\,\ref{resonant-spectra}c, verifying that we also reach a high efficiency in coupling a molecule to single photons. We note in passing that one of the challenges in this experiment concerns tuning the frequency of single photons, which we realized via Stark effect on the ``source" molecule.

In future, the degree of mastery demonstrated here can be combined with pulsed excitation of the source molecule and extended to the coupling of two or more photons to a single molecule\,\cite{Chang:14,Maser:16}. Such an experiment establishes a platform for nonlinear quantum optics at its most fundamental level for the realization of gates and for quantum information processing\,\cite{Kok2010}. 

\section{Discussion and outlook}
The large phase shift, nonlinearity at the single-photon level and strongly nonclassical photon statistics demonstrated in this work give access to a range of quantum functionalities such as photon sorting and gates\,\cite{Kok2010,Ralph:15} in organic materials. In a next step, chip-based ring resonators\,\cite{Rotenberg:17} and nanoguides\,\cite{Tuerschmann:17} will apply these opportunities to nanophotonic circuits. Having reached a highly efficient level of interaction between single photons and single molecules, one can then devise novel linear and nonlinear cooperative effects and polaritonic states, where a controlled number of molecules and photons are coupled via a common photonic mode along the circuit \,\cite{Diniz:11,Haakh:16}. The practical implementation of these concepts would particularly benefit from the use of polymer media instead of organic crystals for device fabrication. Indeed, the Purcell enhanced radiative rates achieved in our work already compete with and dominate phonon-induced dephasing rates in polymers, which lie in the range of 0.1-1 GHz\,\cite{Walser:09}. In addition to their immediate potential for large-scale organic quantum networks, we expect the selective modification of molecular rates demonstrated in this work to find applications in cooling and trapping of molecules in the gas phase, where closed transitions are desirable \,\cite{Kozyryev:17} and in control of molecular photochemical processes such as photochromic switching\,\cite{Schwartz:11}.

\section*{Acknowledgments}
We thank Jan Renger, Anke Dutschke, and Eduard Butzen for the fabrication of micromirrors, Maksim Schwab for the construction of the mechanical components for the cryostats, and Lothar Meier for help with electronics to lock the cavity. This work was supported by the Max Planck Society.

\bibliographystyle{aipnum4-1}

\begin{thebibliography}{46}%
\makeatletter
\providecommand \@ifxundefined [1]{%
 \@ifx{#1\undefined}
}%
\providecommand \@ifnum [1]{%
 \ifnum #1\expandafter \@firstoftwo
 \else \expandafter \@secondoftwo
 \fi
}%
\providecommand \@ifx [1]{%
 \ifx #1\expandafter \@firstoftwo
 \else \expandafter \@secondoftwo
 \fi
}%
\providecommand \natexlab [1]{#1}%
\providecommand \enquote  [1]{``#1''}%
\providecommand \bibnamefont  [1]{#1}%
\providecommand \bibfnamefont [1]{#1}%
\providecommand \citenamefont [1]{#1}%
\providecommand \href@noop [0]{\@secondoftwo}%
\providecommand \href [0]{\begingroup \@sanitize@url \@href}%
\providecommand \@href[1]{\@@startlink{#1}\@@href}%
\providecommand \@@href[1]{\endgroup#1\@@endlink}%
\providecommand \@sanitize@url [0]{\catcode `\\12\catcode `\$12\catcode
  `\&12\catcode `\#12\catcode `\^12\catcode `\_12\catcode `\%12\relax}%
\providecommand \@@startlink[1]{}%
\providecommand \@@endlink[0]{}%
\providecommand \url  [0]{\begingroup\@sanitize@url \@url }%
\providecommand \@url [1]{\endgroup\@href {#1}{\urlprefix }}%
\providecommand \urlprefix  [0]{URL }%
\providecommand \Eprint [0]{\href }%
\providecommand \doibase [0]{http://dx.doi.org/}%
\providecommand \selectlanguage [0]{\@gobble}%
\providecommand \bibinfo  [0]{\@secondoftwo}%
\providecommand \bibfield  [0]{\@secondoftwo}%
\providecommand \translation [1]{[#1]}%
\providecommand \BibitemOpen [0]{}%
\providecommand \bibitemStop [0]{}%
\providecommand \bibitemNoStop [0]{.\EOS\space}%
\providecommand \EOS [0]{\spacefactor3000\relax}%
\providecommand \BibitemShut  [1]{\csname bibitem#1\endcsname}%
\let\auto@bib@innerbib\@empty
\bibitem [{\citenamefont {Spaun}\ \emph {et~al.}(2016)\citenamefont {Spaun},
  \citenamefont {Changala}, \citenamefont {Patterson}, \citenamefont {Bjork},
  \citenamefont {Heckl}, \citenamefont {Doyle},\ and\ \citenamefont
  {Ye}}]{Spaun:16}%
  \BibitemOpen
  \bibfield  {author} {\bibinfo {author} {\bibfnamefont {B.}~\bibnamefont
  {Spaun}}, \bibinfo {author} {\bibfnamefont {P.~B.}\ \bibnamefont {Changala}},
  \bibinfo {author} {\bibfnamefont {D.}~\bibnamefont {Patterson}}, \bibinfo
  {author} {\bibfnamefont {B.~J.}\ \bibnamefont {Bjork}}, \bibinfo {author}
  {\bibfnamefont {O.~H.}\ \bibnamefont {Heckl}}, \bibinfo {author}
  {\bibfnamefont {J.~M.}\ \bibnamefont {Doyle}}, \ and\ \bibinfo {author}
  {\bibfnamefont {J.}~\bibnamefont {Ye}},\ }\href@noop {} {\bibfield  {journal}
  {\bibinfo  {journal} {Nature}\ }\textbf {\bibinfo {volume} {533}},\ \bibinfo
  {pages} {517} (\bibinfo {year} {2016})}\BibitemShut {NoStop}%
\bibitem [{\citenamefont {Truppe}\ \emph {et~al.}(2017)\citenamefont {Truppe},
  \citenamefont {Williams}, \citenamefont {Hambach}, \citenamefont {Caldwell},
  \citenamefont {Fitch}, \citenamefont {Hinds}, \citenamefont {Sauer},\ and\
  \citenamefont {Tarbutt}}]{Truppe:17}%
  \BibitemOpen
  \bibfield  {author} {\bibinfo {author} {\bibfnamefont {S.}~\bibnamefont
  {Truppe}}, \bibinfo {author} {\bibfnamefont {H.~J.}\ \bibnamefont
  {Williams}}, \bibinfo {author} {\bibfnamefont {M.}~\bibnamefont {Hambach}},
  \bibinfo {author} {\bibfnamefont {L.}~\bibnamefont {Caldwell}}, \bibinfo
  {author} {\bibfnamefont {N.~J.}\ \bibnamefont {Fitch}}, \bibinfo {author}
  {\bibfnamefont {E.~A.}\ \bibnamefont {Hinds}}, \bibinfo {author}
  {\bibfnamefont {B.~E.}\ \bibnamefont {Sauer}}, \ and\ \bibinfo {author}
  {\bibfnamefont {M.~R.}\ \bibnamefont {Tarbutt}},\ }\href@noop {} {\bibfield
  {journal} {\bibinfo  {journal} {Nat. Phys.}\ }\textbf {\bibinfo {volume}
  {13}},\ \bibinfo {pages} {1173} (\bibinfo {year} {2017})}\BibitemShut
  {NoStop}%
\bibitem [{\citenamefont {Lidzey}\ \emph {et~al.}(2000)\citenamefont {Lidzey},
  \citenamefont {Bradley}, \citenamefont {Armitage}, \citenamefont {Walker},\
  and\ \citenamefont {Skolnick}}]{Lidzey:00}%
  \BibitemOpen
  \bibfield  {author} {\bibinfo {author} {\bibfnamefont {D.~G.}\ \bibnamefont
  {Lidzey}}, \bibinfo {author} {\bibfnamefont {D.~D.~C.}\ \bibnamefont
  {Bradley}}, \bibinfo {author} {\bibfnamefont {A.}~\bibnamefont {Armitage}},
  \bibinfo {author} {\bibfnamefont {S.}~\bibnamefont {Walker}}, \ and\ \bibinfo
  {author} {\bibfnamefont {M.~S.}\ \bibnamefont {Skolnick}},\ }\href@noop {}
  {\bibfield  {journal} {\bibinfo  {journal} {Science}\ }\textbf {\bibinfo
  {volume} {288}},\ \bibinfo {pages} {1620} (\bibinfo {year}
  {2000})}\BibitemShut {NoStop}%
\bibitem [{\citenamefont {Hakala}\ \emph {et~al.}(2009)\citenamefont {Hakala},
  \citenamefont {Toppari}, \citenamefont {Kuzyk}, \citenamefont {Pettersson},
  \citenamefont {Tikkanen}, \citenamefont {Kunttu},\ and\ \citenamefont
  {T\"orm\"a}}]{Hakala:09}%
  \BibitemOpen
  \bibfield  {author} {\bibinfo {author} {\bibfnamefont {T.~K.}\ \bibnamefont
  {Hakala}}, \bibinfo {author} {\bibfnamefont {J.~J.}\ \bibnamefont {Toppari}},
  \bibinfo {author} {\bibfnamefont {A.}~\bibnamefont {Kuzyk}}, \bibinfo
  {author} {\bibfnamefont {M.}~\bibnamefont {Pettersson}}, \bibinfo {author}
  {\bibfnamefont {H.}~\bibnamefont {Tikkanen}}, \bibinfo {author}
  {\bibfnamefont {H.}~\bibnamefont {Kunttu}}, \ and\ \bibinfo {author}
  {\bibfnamefont {P.}~\bibnamefont {T\"orm\"a}},\ }\href@noop {} {\bibfield
  {journal} {\bibinfo  {journal} {Phys. Rev. Lett.}\ }\textbf {\bibinfo
  {volume} {103}},\ \bibinfo {pages} {053602} (\bibinfo {year}
  {2009})}\BibitemShut {NoStop}%
\bibitem [{\citenamefont {Schwartz}\ \emph {et~al.}(2011)\citenamefont
  {Schwartz}, \citenamefont {Hutchison}, \citenamefont {Genet},\ and\
  \citenamefont {Ebbesen}}]{Schwartz:11}%
  \BibitemOpen
  \bibfield  {author} {\bibinfo {author} {\bibfnamefont {T.}~\bibnamefont
  {Schwartz}}, \bibinfo {author} {\bibfnamefont {J.~A.}\ \bibnamefont
  {Hutchison}}, \bibinfo {author} {\bibfnamefont {C.}~\bibnamefont {Genet}}, \
  and\ \bibinfo {author} {\bibfnamefont {T.~W.}\ \bibnamefont {Ebbesen}},\
  }\href@noop {} {\bibfield  {journal} {\bibinfo  {journal} {Phys. Rev. Lett.}\
  }\textbf {\bibinfo {volume} {106}},\ \bibinfo {pages} {196405} (\bibinfo
  {year} {2011})}\BibitemShut {NoStop}%
\bibitem [{\citenamefont {Chikkaraddy}\ \emph {et~al.}(2016)\citenamefont
  {Chikkaraddy}, \citenamefont {de~Nijs}, \citenamefont {Benz}, \citenamefont
  {Barrow}, \citenamefont {Scherman}, \citenamefont {Rosta}, \citenamefont
  {Demetriadou}, \citenamefont {Fox}, \citenamefont {Hess},\ and\ \citenamefont
  {Baumberg}}]{Chikkaraddy:16}%
  \BibitemOpen
  \bibfield  {author} {\bibinfo {author} {\bibfnamefont {R.}~\bibnamefont
  {Chikkaraddy}}, \bibinfo {author} {\bibfnamefont {B.}~\bibnamefont
  {de~Nijs}}, \bibinfo {author} {\bibfnamefont {F.}~\bibnamefont {Benz}},
  \bibinfo {author} {\bibfnamefont {S.~J.}\ \bibnamefont {Barrow}}, \bibinfo
  {author} {\bibfnamefont {O.~A.}\ \bibnamefont {Scherman}}, \bibinfo {author}
  {\bibfnamefont {E.}~\bibnamefont {Rosta}}, \bibinfo {author} {\bibfnamefont
  {A.}~\bibnamefont {Demetriadou}}, \bibinfo {author} {\bibfnamefont
  {P.}~\bibnamefont {Fox}}, \bibinfo {author} {\bibfnamefont {O.}~\bibnamefont
  {Hess}}, \ and\ \bibinfo {author} {\bibfnamefont {J.~J.}\ \bibnamefont
  {Baumberg}},\ }\href@noop {} {\bibfield  {journal} {\bibinfo  {journal}
  {Nature}\ }\textbf {\bibinfo {volume} {535}},\ \bibinfo {pages} {127}
  (\bibinfo {year} {2016})}\BibitemShut {NoStop}%
\bibitem [{\citenamefont {Polisseni}\ \emph {et~al.}(2016)\citenamefont
  {Polisseni}, \citenamefont {Major}, \citenamefont {Boissier}, \citenamefont
  {Grandi}, \citenamefont {Clark},\ and\ \citenamefont {Hinds}}]{Polisseni:16}%
  \BibitemOpen
  \bibfield  {author} {\bibinfo {author} {\bibfnamefont {C.}~\bibnamefont
  {Polisseni}}, \bibinfo {author} {\bibfnamefont {K.~D.}\ \bibnamefont
  {Major}}, \bibinfo {author} {\bibfnamefont {S.}~\bibnamefont {Boissier}},
  \bibinfo {author} {\bibfnamefont {S.}~\bibnamefont {Grandi}}, \bibinfo
  {author} {\bibfnamefont {A.~S.}\ \bibnamefont {Clark}}, \ and\ \bibinfo
  {author} {\bibfnamefont {E.~A.}\ \bibnamefont {Hinds}},\ }\href@noop {}
  {\bibfield  {journal} {\bibinfo  {journal} {Opt. Express}\ }\textbf {\bibinfo
  {volume} {24}},\ \bibinfo {pages} {5615} (\bibinfo {year}
  {2016})}\BibitemShut {NoStop}%
\bibitem [{\citenamefont {Zhang}\ \emph {et~al.}(2017)\citenamefont {Zhang},
  \citenamefont {Meng}, \citenamefont {Zhang}, \citenamefont {Luo},
  \citenamefont {Yu}, \citenamefont {Yang}, \citenamefont {Zhang},
  \citenamefont {Esteban}, \citenamefont {Aizpurua}, \citenamefont {Luo},\ and\
  \citenamefont {et~al.}}]{Zhang:17}%
  \BibitemOpen
  \bibfield  {author} {\bibinfo {author} {\bibfnamefont {Y.}~\bibnamefont
  {Zhang}}, \bibinfo {author} {\bibfnamefont {Q.-S.}\ \bibnamefont {Meng}},
  \bibinfo {author} {\bibfnamefont {L.}~\bibnamefont {Zhang}}, \bibinfo
  {author} {\bibfnamefont {Y.}~\bibnamefont {Luo}}, \bibinfo {author}
  {\bibfnamefont {Y.-J.}\ \bibnamefont {Yu}}, \bibinfo {author} {\bibfnamefont
  {B.}~\bibnamefont {Yang}}, \bibinfo {author} {\bibfnamefont {Y.}~\bibnamefont
  {Zhang}}, \bibinfo {author} {\bibfnamefont {R.}~\bibnamefont {Esteban}},
  \bibinfo {author} {\bibfnamefont {J.}~\bibnamefont {Aizpurua}}, \bibinfo
  {author} {\bibfnamefont {Y.}~\bibnamefont {Luo}}, \ and\ \bibinfo {author}
  {\bibnamefont {et~al.}},\ }\href@noop {} {\bibfield  {journal} {\bibinfo
  {journal} {Nat. Commun.}\ }\textbf {\bibinfo {volume} {8}},\ \bibinfo {pages}
  {15225} (\bibinfo {year} {2017})}\BibitemShut {NoStop}%
\bibitem [{\citenamefont {Skoff}\ \emph {et~al.}(2018)\citenamefont {Skoff},
  \citenamefont {Papencordt}, \citenamefont {Schauffert}, \citenamefont
  {Bayer},\ and\ \citenamefont {Rauschenbeutel}}]{Skoff:18}%
  \BibitemOpen
  \bibfield  {author} {\bibinfo {author} {\bibfnamefont {S.~M.}\ \bibnamefont
  {Skoff}}, \bibinfo {author} {\bibfnamefont {D.}~\bibnamefont {Papencordt}},
  \bibinfo {author} {\bibfnamefont {H.}~\bibnamefont {Schauffert}}, \bibinfo
  {author} {\bibfnamefont {B.~C.}\ \bibnamefont {Bayer}}, \ and\ \bibinfo
  {author} {\bibfnamefont {A.}~\bibnamefont {Rauschenbeutel}},\ }\href@noop {}
  {\bibfield  {journal} {\bibinfo  {journal} {Phys. Rev. A}\ }\textbf {\bibinfo
  {volume} {97}},\ \bibinfo {pages} {043839} (\bibinfo {year}
  {2018})}\BibitemShut {NoStop}%
\bibitem [{\citenamefont {Liu}\ \emph {et~al.}(2018)\citenamefont {Liu},
  \citenamefont {Hood}, \citenamefont {Yu}, \citenamefont {Zhang},
  \citenamefont {Hutzler}, \citenamefont {Rosenband},\ and\ \citenamefont
  {Ni}}]{Liu:18}%
  \BibitemOpen
  \bibfield  {author} {\bibinfo {author} {\bibfnamefont {L.~R.}\ \bibnamefont
  {Liu}}, \bibinfo {author} {\bibfnamefont {J.~D.}\ \bibnamefont {Hood}},
  \bibinfo {author} {\bibfnamefont {Y.}~\bibnamefont {Yu}}, \bibinfo {author}
  {\bibfnamefont {J.~T.}\ \bibnamefont {Zhang}}, \bibinfo {author}
  {\bibfnamefont {N.~R.}\ \bibnamefont {Hutzler}}, \bibinfo {author}
  {\bibfnamefont {T.}~\bibnamefont {Rosenband}}, \ and\ \bibinfo {author}
  {\bibfnamefont {K.-K.}\ \bibnamefont {Ni}},\ }\href@noop {} {\bibfield
  {journal} {\bibinfo  {journal} {Science}\ }\textbf {\bibinfo {volume}
  {360}},\ \bibinfo {pages} {900} (\bibinfo {year} {2018})}\BibitemShut
  {NoStop}%
\bibitem [{\citenamefont {Moerner}\ \emph {et~al.}(1996)\citenamefont
  {Moerner}, \citenamefont {Orrit}, \citenamefont {Wild},\ and\ \citenamefont
  {Basch{\'e}}}]{SMbook}%
  \BibitemOpen
  \bibinfo {editor} {\bibfnamefont {W.}~\bibnamefont {Moerner}}, \bibinfo
  {editor} {\bibfnamefont {M.}~\bibnamefont {Orrit}}, \bibinfo {editor}
  {\bibfnamefont {U.}~\bibnamefont {Wild}}, \ and\ \bibinfo {editor}
  {\bibfnamefont {T.}~\bibnamefont {Basch{\'e}}},\ eds.,\ \href@noop {} {\emph
  {\bibinfo {title} {Single-Molecule Optical Detection, Imaging and
  Spectroscopy}}}\ (\bibinfo  {publisher} {Wiley},\ \bibinfo {year}
  {1996})\BibitemShut {NoStop}%
\bibitem [{\citenamefont {Pototschnig}\ \emph {et~al.}(2011)\citenamefont
  {Pototschnig}, \citenamefont {Chassagneux}, \citenamefont {Hwang},
  \citenamefont {Zumofen}, \citenamefont {Renn},\ and\ \citenamefont
  {Sandoghdar}}]{Pototschnig:11}%
  \BibitemOpen
  \bibfield  {author} {\bibinfo {author} {\bibfnamefont {M.}~\bibnamefont
  {Pototschnig}}, \bibinfo {author} {\bibfnamefont {Y.}~\bibnamefont
  {Chassagneux}}, \bibinfo {author} {\bibfnamefont {J.}~\bibnamefont {Hwang}},
  \bibinfo {author} {\bibfnamefont {G.}~\bibnamefont {Zumofen}}, \bibinfo
  {author} {\bibfnamefont {A.}~\bibnamefont {Renn}}, \ and\ \bibinfo {author}
  {\bibfnamefont {V.}~\bibnamefont {Sandoghdar}},\ }\href@noop {} {\bibfield
  {journal} {\bibinfo  {journal} {Phys. Rev. Lett.}\ }\textbf {\bibinfo
  {volume} {107}},\ \bibinfo {pages} {063001} (\bibinfo {year}
  {2011})}\BibitemShut {NoStop}%
\bibitem [{\citenamefont {Maser}\ \emph {et~al.}(2016)\citenamefont {Maser},
  \citenamefont {Gmeiner}, \citenamefont {Utikal}, \citenamefont
  {G{\"o}tzinger},\ and\ \citenamefont {Sandoghdar}}]{Maser:16}%
  \BibitemOpen
  \bibfield  {author} {\bibinfo {author} {\bibfnamefont {A.}~\bibnamefont
  {Maser}}, \bibinfo {author} {\bibfnamefont {B.}~\bibnamefont {Gmeiner}},
  \bibinfo {author} {\bibfnamefont {T.}~\bibnamefont {Utikal}}, \bibinfo
  {author} {\bibfnamefont {S.}~\bibnamefont {G{\"o}tzinger}}, \ and\ \bibinfo
  {author} {\bibfnamefont {V.}~\bibnamefont {Sandoghdar}},\ }\href@noop {}
  {\bibfield  {journal} {\bibinfo  {journal} {Nat. Photon.}\ }\textbf {\bibinfo
  {volume} {10}},\ \bibinfo {pages} {450} (\bibinfo {year} {2016})}\BibitemShut
  {NoStop}%
\bibitem [{\citenamefont {K\"uhn}\ \emph {et~al.}(2006)\citenamefont {K\"uhn},
  \citenamefont {{H\aa kanson}}, \citenamefont {Rogobete},\ and\ \citenamefont
  {Sandoghdar}}]{Kuehn:06}%
  \BibitemOpen
  \bibfield  {author} {\bibinfo {author} {\bibfnamefont {S.}~\bibnamefont
  {K\"uhn}}, \bibinfo {author} {\bibfnamefont {U.}~\bibnamefont {{H\aa
  kanson}}}, \bibinfo {author} {\bibfnamefont {L.}~\bibnamefont {Rogobete}}, \
  and\ \bibinfo {author} {\bibfnamefont {V.}~\bibnamefont {Sandoghdar}},\
  }\href@noop {} {\bibfield  {journal} {\bibinfo  {journal} {Phys. Rev. Lett.}\
  }\textbf {\bibinfo {volume} {97}},\ \bibinfo {eid} {017402} (\bibinfo {year}
  {2006})}\BibitemShut {NoStop}%
\bibitem [{\citenamefont {Berman}(1994)}]{book-Berman94}%
  \BibitemOpen
  \bibinfo {editor} {\bibfnamefont {P.~R.}\ \bibnamefont {Berman}},\ ed.,\
  \href@noop {} {\emph {\bibinfo {title} {Cavity quantum electrodynamics}}}\
  (\bibinfo  {publisher} {Academic Press},\ \bibinfo {address} {San Diego},\
  \bibinfo {year} {1994})\BibitemShut {NoStop}%
\bibitem [{\citenamefont {Norris}, \citenamefont {Kuwata-Gonokami},\ and\
  \citenamefont {Moerner}(1997)}]{Norris:97}%
  \BibitemOpen
  \bibfield  {author} {\bibinfo {author} {\bibfnamefont {D.~J.}\ \bibnamefont
  {Norris}}, \bibinfo {author} {\bibfnamefont {M.}~\bibnamefont
  {Kuwata-Gonokami}}, \ and\ \bibinfo {author} {\bibfnamefont {W.~E.}\
  \bibnamefont {Moerner}},\ }\href@noop {} {\bibfield  {journal} {\bibinfo
  {journal} {Appl. Phys. Lett.}\ }\textbf {\bibinfo {volume} {71}},\ \bibinfo
  {pages} {297} (\bibinfo {year} {1997})}\BibitemShut {NoStop}%
\bibitem [{\citenamefont {Steiner}\ \emph {et~al.}(2007)\citenamefont
  {Steiner}, \citenamefont {Hartschuh}, \citenamefont {Korlacki},\ and\
  \citenamefont {Meixner}}]{Steiner:07}%
  \BibitemOpen
  \bibfield  {author} {\bibinfo {author} {\bibfnamefont {M.}~\bibnamefont
  {Steiner}}, \bibinfo {author} {\bibfnamefont {A.}~\bibnamefont {Hartschuh}},
  \bibinfo {author} {\bibfnamefont {R.}~\bibnamefont {Korlacki}}, \ and\
  \bibinfo {author} {\bibfnamefont {A.~J.}\ \bibnamefont {Meixner}},\
  }\href@noop {} {\bibfield  {journal} {\bibinfo  {journal} {Appl. Phys.
  Lett.}\ }\textbf {\bibinfo {volume} {90}},\ \bibinfo {pages} {183122}
  (\bibinfo {year} {2007})}\BibitemShut {NoStop}%
\bibitem [{\citenamefont {Chizhik}\ \emph {et~al.}(2009)\citenamefont
  {Chizhik}, \citenamefont {Schleifenbaum}, \citenamefont {Gutbrod},
  \citenamefont {Chizhik}, \citenamefont {Khoptyar}, \citenamefont {Meixner},\
  and\ \citenamefont {Enderlein}}]{Chizhik:09}%
  \BibitemOpen
  \bibfield  {author} {\bibinfo {author} {\bibfnamefont {A.}~\bibnamefont
  {Chizhik}}, \bibinfo {author} {\bibfnamefont {F.}~\bibnamefont
  {Schleifenbaum}}, \bibinfo {author} {\bibfnamefont {R.}~\bibnamefont
  {Gutbrod}}, \bibinfo {author} {\bibfnamefont {A.}~\bibnamefont {Chizhik}},
  \bibinfo {author} {\bibfnamefont {D.}~\bibnamefont {Khoptyar}}, \bibinfo
  {author} {\bibfnamefont {A.~J.}\ \bibnamefont {Meixner}}, \ and\ \bibinfo
  {author} {\bibfnamefont {J.}~\bibnamefont {Enderlein}},\ }\href@noop {}
  {\bibfield  {journal} {\bibinfo  {journal} {Phys. Rev. Lett.}\ }\textbf
  {\bibinfo {volume} {102}},\ \bibinfo {pages} {073002} (\bibinfo {year}
  {2009})}\BibitemShut {NoStop}%
\bibitem [{\citenamefont {Wang}\ \emph {et~al.}(2017)\citenamefont {Wang},
  \citenamefont {Kelkar}, \citenamefont {Martin-Cano}, \citenamefont {Utikal},
  \citenamefont {G{\"{o}}tzinger},\ and\ \citenamefont {Sandoghdar}}]{Wang:17}%
  \BibitemOpen
  \bibfield  {author} {\bibinfo {author} {\bibfnamefont {D.}~\bibnamefont
  {Wang}}, \bibinfo {author} {\bibfnamefont {H.}~\bibnamefont {Kelkar}},
  \bibinfo {author} {\bibfnamefont {D.}~\bibnamefont {Martin-Cano}}, \bibinfo
  {author} {\bibfnamefont {T.}~\bibnamefont {Utikal}}, \bibinfo {author}
  {\bibfnamefont {S.}~\bibnamefont {G{\"{o}}tzinger}}, \ and\ \bibinfo {author}
  {\bibfnamefont {V.}~\bibnamefont {Sandoghdar}},\ }\href@noop {} {\bibfield
  {journal} {\bibinfo  {journal} {Phys. Rev. X}\ }\textbf {\bibinfo {volume}
  {7}},\ \bibinfo {pages} {021014} (\bibinfo {year} {2017})}\BibitemShut
  {NoStop}%
\bibitem [{\citenamefont {Toninelli}\ \emph {et~al.}(2010)\citenamefont
  {Toninelli}, \citenamefont {Delley}, \citenamefont {St{\"o}ferle},
  \citenamefont {Renn}, \citenamefont {G{\"o}tzinger},\ and\ \citenamefont
  {Sandoghdar}}]{Toninelli:10}%
  \BibitemOpen
  \bibfield  {author} {\bibinfo {author} {\bibfnamefont {C.}~\bibnamefont
  {Toninelli}}, \bibinfo {author} {\bibfnamefont {Y.}~\bibnamefont {Delley}},
  \bibinfo {author} {\bibfnamefont {T.}~\bibnamefont {St{\"o}ferle}}, \bibinfo
  {author} {\bibfnamefont {A.}~\bibnamefont {Renn}}, \bibinfo {author}
  {\bibfnamefont {S.}~\bibnamefont {G{\"o}tzinger}}, \ and\ \bibinfo {author}
  {\bibfnamefont {V.}~\bibnamefont {Sandoghdar}},\ }\href@noop {} {\bibfield
  {journal} {\bibinfo  {journal} {Appl. Phys. Lett.}\ }\textbf {\bibinfo
  {volume} {97}},\ \bibinfo {pages} {021107} (\bibinfo {year}
  {2010})}\BibitemShut {NoStop}%
\bibitem [{\citenamefont {Kelkar}\ \emph {et~al.}(2015)\citenamefont {Kelkar},
  \citenamefont {Wang}, \citenamefont {Mart{\'\i}n-Cano}, \citenamefont
  {Hoffmann}, \citenamefont {Christiansen}, \citenamefont {G{\"o}tzinger},\
  and\ \citenamefont {Sandoghdar}}]{Kelkar:15}%
  \BibitemOpen
  \bibfield  {author} {\bibinfo {author} {\bibfnamefont {H.}~\bibnamefont
  {Kelkar}}, \bibinfo {author} {\bibfnamefont {D.}~\bibnamefont {Wang}},
  \bibinfo {author} {\bibfnamefont {D.}~\bibnamefont {Mart{\'\i}n-Cano}},
  \bibinfo {author} {\bibfnamefont {B.}~\bibnamefont {Hoffmann}}, \bibinfo
  {author} {\bibfnamefont {S.}~\bibnamefont {Christiansen}}, \bibinfo {author}
  {\bibfnamefont {S.}~\bibnamefont {G{\"o}tzinger}}, \ and\ \bibinfo {author}
  {\bibfnamefont {V.}~\bibnamefont {Sandoghdar}},\ }\href@noop {} {\bibfield
  {journal} {\bibinfo  {journal} {Phys. Rev. Appl.}\ }\textbf {\bibinfo
  {volume} {4}},\ \bibinfo {pages} {054010} (\bibinfo {year}
  {2015})}\BibitemShut {NoStop}%
\bibitem [{\citenamefont {Nicolet}\ \emph {et~al.}(2007)\citenamefont
  {Nicolet}, \citenamefont {Hofmann}, \citenamefont {Kol'chenko}, \citenamefont
  {Kozankiewicz},\ and\ \citenamefont {Orrit}}]{Nicolet:07}%
  \BibitemOpen
  \bibfield  {author} {\bibinfo {author} {\bibfnamefont {A.~A.~L.}\
  \bibnamefont {Nicolet}}, \bibinfo {author} {\bibfnamefont {C.}~\bibnamefont
  {Hofmann}}, \bibinfo {author} {\bibfnamefont {M.~A.}\ \bibnamefont
  {Kol'chenko}}, \bibinfo {author} {\bibfnamefont {B.}~\bibnamefont
  {Kozankiewicz}}, \ and\ \bibinfo {author} {\bibfnamefont {M.}~\bibnamefont
  {Orrit}},\ }\href@noop {} {\bibfield  {journal} {\bibinfo  {journal}
  {ChemPhysChem}\ }\textbf {\bibinfo {volume} {8}},\ \bibinfo {pages} {1215}
  (\bibinfo {year} {2007})}\BibitemShut {NoStop}%
\bibitem [{\citenamefont {Trebbia}\ \emph {et~al.}(2009)\citenamefont
  {Trebbia}, \citenamefont {Ruf}, \citenamefont {Tamarat},\ and\ \citenamefont
  {Lounis}}]{Trebbia:09}%
  \BibitemOpen
  \bibfield  {author} {\bibinfo {author} {\bibfnamefont {J.-B.}\ \bibnamefont
  {Trebbia}}, \bibinfo {author} {\bibfnamefont {H.}~\bibnamefont {Ruf}},
  \bibinfo {author} {\bibfnamefont {P.}~\bibnamefont {Tamarat}}, \ and\
  \bibinfo {author} {\bibfnamefont {B.}~\bibnamefont {Lounis}},\ }\href@noop {}
  {\bibfield  {journal} {\bibinfo  {journal} {Opt. Express}\ }\textbf {\bibinfo
  {volume} {17}},\ \bibinfo {pages} {23986} (\bibinfo {year}
  {2009})}\BibitemShut {NoStop}%
\bibitem [{\citenamefont {Petermann}(1979)}]{Petermann:79}%
  \BibitemOpen
  \bibfield  {author} {\bibinfo {author} {\bibfnamefont {K.}~\bibnamefont
  {Petermann}},\ }\href@noop {} {\bibfield  {journal} {\bibinfo  {journal}
  {IEEE J. Quantum Electron.}\ }\textbf {\bibinfo {volume} {15}},\ \bibinfo
  {pages} {566} (\bibinfo {year} {1979})}\BibitemShut {NoStop}%
\bibitem [{\citenamefont {Auff{\`{e}}ves-Garnier}\ \emph
  {et~al.}(2007)\citenamefont {Auff{\`{e}}ves-Garnier}, \citenamefont {Simon},
  \citenamefont {G{\'{e}}rard},\ and\ \citenamefont {Poizat}}]{Auffeves:07}%
  \BibitemOpen
  \bibfield  {author} {\bibinfo {author} {\bibfnamefont {A.}~\bibnamefont
  {Auff{\`{e}}ves-Garnier}}, \bibinfo {author} {\bibfnamefont {C.}~\bibnamefont
  {Simon}}, \bibinfo {author} {\bibfnamefont {J.~M.}\ \bibnamefont
  {G{\'{e}}rard}}, \ and\ \bibinfo {author} {\bibfnamefont {J.~P.}\
  \bibnamefont {Poizat}},\ }\href@noop {} {\bibfield  {journal} {\bibinfo
  {journal} {Phys. Rev. A.}\ }\textbf {\bibinfo {volume} {75}},\ \bibinfo
  {pages} {053823} (\bibinfo {year} {2007})}\BibitemShut {NoStop}%
\bibitem [{\citenamefont {Heinzen}\ and\ \citenamefont
  {Feld}(1987)}]{Heinzen:87}%
  \BibitemOpen
  \bibfield  {author} {\bibinfo {author} {\bibfnamefont {D.~J.}\ \bibnamefont
  {Heinzen}}\ and\ \bibinfo {author} {\bibfnamefont {M.~S.}\ \bibnamefont
  {Feld}},\ }\href@noop {} {\bibfield  {journal} {\bibinfo  {journal} {Phys.
  Rev. Lett.}\ }\textbf {\bibinfo {volume} {59}},\ \bibinfo {pages} {2623}
  (\bibinfo {year} {1987})}\BibitemShut {NoStop}%
\bibitem [{\citenamefont {Cohen-Tannoudji}, \citenamefont {Dupont-Roc},\ and\
  \citenamefont {Grynberg}(2004)}]{cohen-book}%
  \BibitemOpen
  \bibfield  {author} {\bibinfo {author} {\bibfnamefont {C.}~\bibnamefont
  {Cohen-Tannoudji}}, \bibinfo {author} {\bibfnamefont {J.}~\bibnamefont
  {Dupont-Roc}}, \ and\ \bibinfo {author} {\bibfnamefont {G.}~\bibnamefont
  {Grynberg}},\ }\href@noop {} {\emph {\bibinfo {title} {Atom-Photon
  Interactions}}}\ (\bibinfo  {publisher} {Wiley-VCH},\ \bibinfo {address}
  {Weinheim, Germany},\ \bibinfo {year} {2004})\BibitemShut {NoStop}%
\bibitem [{\citenamefont {Choi}\ \emph {et~al.}(2010)\citenamefont {Choi},
  \citenamefont {Kang}, \citenamefont {Lim}, \citenamefont {Kim}, \citenamefont
  {Kim}, \citenamefont {Lee},\ and\ \citenamefont {An}}]{Choi:10}%
  \BibitemOpen
  \bibfield  {author} {\bibinfo {author} {\bibfnamefont {Y.}~\bibnamefont
  {Choi}}, \bibinfo {author} {\bibfnamefont {S.}~\bibnamefont {Kang}}, \bibinfo
  {author} {\bibfnamefont {S.}~\bibnamefont {Lim}}, \bibinfo {author}
  {\bibfnamefont {W.}~\bibnamefont {Kim}}, \bibinfo {author} {\bibfnamefont
  {J.-R.}\ \bibnamefont {Kim}}, \bibinfo {author} {\bibfnamefont {J.-H.}\
  \bibnamefont {Lee}}, \ and\ \bibinfo {author} {\bibfnamefont
  {K.}~\bibnamefont {An}},\ }\href@noop {} {\bibfield  {journal} {\bibinfo
  {journal} {Phys. Rev. Lett.}\ }\textbf {\bibinfo {volume} {104}},\ \bibinfo
  {pages} {153601} (\bibinfo {year} {2010})}\BibitemShut {NoStop}%
\bibitem [{\citenamefont {Rice}\ and\ \citenamefont
  {Carmichael}(1988)}]{Rice:88}%
  \BibitemOpen
  \bibfield  {author} {\bibinfo {author} {\bibfnamefont {P.~R.}\ \bibnamefont
  {Rice}}\ and\ \bibinfo {author} {\bibfnamefont {H.~J.}\ \bibnamefont
  {Carmichael}},\ }\href@noop {} {\bibfield  {journal} {\bibinfo  {journal}
  {IEEE J. Quantum Electron.}\ }\textbf {\bibinfo {volume} {24}},\ \bibinfo
  {pages} {1351} (\bibinfo {year} {1988})}\BibitemShut {NoStop}%
\bibitem [{\citenamefont {Ourjoumtsev}\ \emph {et~al.}(2011)\citenamefont
  {Ourjoumtsev}, \citenamefont {Kubanek}, \citenamefont {Koch}, \citenamefont
  {Sames}, \citenamefont {Pinkse}, \citenamefont {Rempe},\ and\ \citenamefont
  {Murr}}]{Ourjoumtsev2011}%
  \BibitemOpen
  \bibfield  {author} {\bibinfo {author} {\bibfnamefont {A.}~\bibnamefont
  {Ourjoumtsev}}, \bibinfo {author} {\bibfnamefont {A.}~\bibnamefont
  {Kubanek}}, \bibinfo {author} {\bibfnamefont {M.}~\bibnamefont {Koch}},
  \bibinfo {author} {\bibfnamefont {C.}~\bibnamefont {Sames}}, \bibinfo
  {author} {\bibfnamefont {P.~W.~H.}\ \bibnamefont {Pinkse}}, \bibinfo {author}
  {\bibfnamefont {G.}~\bibnamefont {Rempe}}, \ and\ \bibinfo {author}
  {\bibfnamefont {K.}~\bibnamefont {Murr}},\ }\href@noop {} {\bibfield
  {journal} {\bibinfo  {journal} {Nature}\ }\textbf {\bibinfo {volume} {474}},\
  \bibinfo {pages} {623} (\bibinfo {year} {2011})}\BibitemShut {NoStop}%
\bibitem [{\citenamefont {Schulte}\ \emph {et~al.}(2015)\citenamefont
  {Schulte}, \citenamefont {Hansom}, \citenamefont {Jones}, \citenamefont
  {Matthiesen}, \citenamefont {Gall},\ and\ \citenamefont
  {Atature}}]{Schulte2015}%
  \BibitemOpen
  \bibfield  {author} {\bibinfo {author} {\bibfnamefont {C.~H.~H.}\
  \bibnamefont {Schulte}}, \bibinfo {author} {\bibfnamefont {J.}~\bibnamefont
  {Hansom}}, \bibinfo {author} {\bibfnamefont {A.~E.}\ \bibnamefont {Jones}},
  \bibinfo {author} {\bibfnamefont {C.}~\bibnamefont {Matthiesen}}, \bibinfo
  {author} {\bibfnamefont {C.~L.}\ \bibnamefont {Gall}}, \ and\ \bibinfo
  {author} {\bibfnamefont {M.}~\bibnamefont {Atature}},\ }\href@noop {}
  {\bibfield  {journal} {\bibinfo  {journal} {Nature}\ }\textbf {\bibinfo
  {volume} {525}},\ \bibinfo {pages} {222} (\bibinfo {year}
  {2015})}\BibitemShut {NoStop}%
\bibitem [{\citenamefont {Ficek}\ and\ \citenamefont
  {Swain}(2005)}]{book-Ficek}%
  \BibitemOpen
  \bibfield  {author} {\bibinfo {author} {\bibfnamefont {Z.}~\bibnamefont
  {Ficek}}\ and\ \bibinfo {author} {\bibfnamefont {S.}~\bibnamefont {Swain}},\
  }\href@noop {} {\emph {\bibinfo {title} {Quantum Interference and Coherence:
  Theory and Experiments}}}\ (\bibinfo  {publisher} {Springer Science+Business
  Media, Inc},\ \bibinfo {year} {2005})\BibitemShut {NoStop}%
\bibitem [{\citenamefont {Carmichael}\ and\ \citenamefont
  {Rice}(1991)}]{Carmichael:91}%
  \BibitemOpen
  \bibfield  {author} {\bibinfo {author} {\bibfnamefont {H.~J.}\ \bibnamefont
  {Carmichael}}\ and\ \bibinfo {author} {\bibfnamefont {P.~R.}\ \bibnamefont
  {Rice}},\ }\href@noop {} {\bibfield  {journal} {\bibinfo  {journal} {Opt.
  Commun.}\ }\textbf {\bibinfo {volume} {82}},\ \bibinfo {pages} {73} (\bibinfo
  {year} {1991})}\BibitemShut {NoStop}%
\bibitem [{\citenamefont {Bennett}\ \emph {et~al.}(2016)\citenamefont
  {Bennett}, \citenamefont {Lee}, \citenamefont {Ellis}, \citenamefont
  {Farrer}, \citenamefont {Ritchie},\ and\ \citenamefont
  {Shields}}]{Bennett:16}%
  \BibitemOpen
  \bibfield  {author} {\bibinfo {author} {\bibfnamefont {A.~J.}\ \bibnamefont
  {Bennett}}, \bibinfo {author} {\bibfnamefont {J.~P.}\ \bibnamefont {Lee}},
  \bibinfo {author} {\bibfnamefont {D.~J.~P.}\ \bibnamefont {Ellis}}, \bibinfo
  {author} {\bibfnamefont {I.}~\bibnamefont {Farrer}}, \bibinfo {author}
  {\bibfnamefont {D.~A.}\ \bibnamefont {Ritchie}}, \ and\ \bibinfo {author}
  {\bibfnamefont {A.~J.}\ \bibnamefont {Shields}},\ }\href@noop {} {\bibfield
  {journal} {\bibinfo  {journal} {Nat. Nanotech.}\ }\textbf {\bibinfo {volume}
  {11}},\ \bibinfo {pages} {857} (\bibinfo {year} {2016})}\BibitemShut
  {NoStop}%
\bibitem [{\citenamefont {Snijders}\ \emph {et~al.}(2016)\citenamefont
  {Snijders}, \citenamefont {Frey}, \citenamefont {Norman}, \citenamefont
  {Bakker}, \citenamefont {Langman}, \citenamefont {Gossard}, \citenamefont
  {Bowers}, \citenamefont {van Exter}, \citenamefont {Bouwmeester},\ and\
  \citenamefont {L{\"o}ffler}}]{Snijders:16}%
  \BibitemOpen
  \bibfield  {author} {\bibinfo {author} {\bibfnamefont {H.}~\bibnamefont
  {Snijders}}, \bibinfo {author} {\bibfnamefont {J.~A.}\ \bibnamefont {Frey}},
  \bibinfo {author} {\bibfnamefont {J.}~\bibnamefont {Norman}}, \bibinfo
  {author} {\bibfnamefont {M.~P.}\ \bibnamefont {Bakker}}, \bibinfo {author}
  {\bibfnamefont {E.~C.}\ \bibnamefont {Langman}}, \bibinfo {author}
  {\bibfnamefont {A.}~\bibnamefont {Gossard}}, \bibinfo {author} {\bibfnamefont
  {J.~E.}\ \bibnamefont {Bowers}}, \bibinfo {author} {\bibfnamefont {M.~P.}\
  \bibnamefont {van Exter}}, \bibinfo {author} {\bibfnamefont {D.}~\bibnamefont
  {Bouwmeester}}, \ and\ \bibinfo {author} {\bibfnamefont {W.}~\bibnamefont
  {L{\"o}ffler}},\ }\href@noop {} {\bibfield  {journal} {\bibinfo  {journal}
  {Nat. Commun.}\ }\textbf {\bibinfo {volume} {7}},\ \bibinfo {pages} {12578}
  (\bibinfo {year} {2016})}\BibitemShut {NoStop}%
\bibitem [{\citenamefont {Wrigge}\ \emph {et~al.}(2008)\citenamefont {Wrigge},
  \citenamefont {Gerhardt}, \citenamefont {Hwang}, \citenamefont {Zumofen},\
  and\ \citenamefont {Sandoghdar}}]{Wrigge:08}%
  \BibitemOpen
  \bibfield  {author} {\bibinfo {author} {\bibfnamefont {G.}~\bibnamefont
  {Wrigge}}, \bibinfo {author} {\bibfnamefont {I.}~\bibnamefont {Gerhardt}},
  \bibinfo {author} {\bibfnamefont {J.}~\bibnamefont {Hwang}}, \bibinfo
  {author} {\bibfnamefont {G.}~\bibnamefont {Zumofen}}, \ and\ \bibinfo
  {author} {\bibfnamefont {V.}~\bibnamefont {Sandoghdar}},\ }\href@noop {}
  {\bibfield  {journal} {\bibinfo  {journal} {Nat. Phys.}\ }\textbf {\bibinfo
  {volume} {4}},\ \bibinfo {pages} {60} (\bibinfo {year} {2008})}\BibitemShut
  {NoStop}%
\bibitem [{\citenamefont {Chang}, \citenamefont {Vuleti{\'c}},\ and\
  \citenamefont {Lukin}(2014)}]{Chang:14}%
  \BibitemOpen
  \bibfield  {author} {\bibinfo {author} {\bibfnamefont {D.~E.}\ \bibnamefont
  {Chang}}, \bibinfo {author} {\bibfnamefont {V.}~\bibnamefont {Vuleti{\'c}}},
  \ and\ \bibinfo {author} {\bibfnamefont {M.~D.}\ \bibnamefont {Lukin}},\
  }\href@noop {} {\bibfield  {journal} {\bibinfo  {journal} {Nat. Photon.}\
  }\textbf {\bibinfo {volume} {8}},\ \bibinfo {pages} {685} (\bibinfo {year}
  {2014})}\BibitemShut {NoStop}%
\bibitem [{\citenamefont {Rezus}\ \emph {et~al.}(2012)\citenamefont {Rezus},
  \citenamefont {Walt}, \citenamefont {Lettow}, \citenamefont {Renn},
  \citenamefont {Zumofen}, \citenamefont {G\"otzinger},\ and\ \citenamefont
  {Sandoghdar}}]{Rezus:12}%
  \BibitemOpen
  \bibfield  {author} {\bibinfo {author} {\bibfnamefont {Y.~L.~A.}\
  \bibnamefont {Rezus}}, \bibinfo {author} {\bibfnamefont {S.~G.}\ \bibnamefont
  {Walt}}, \bibinfo {author} {\bibfnamefont {R.}~\bibnamefont {Lettow}},
  \bibinfo {author} {\bibfnamefont {A.}~\bibnamefont {Renn}}, \bibinfo {author}
  {\bibfnamefont {G.}~\bibnamefont {Zumofen}}, \bibinfo {author} {\bibfnamefont
  {S.}~\bibnamefont {G\"otzinger}}, \ and\ \bibinfo {author} {\bibfnamefont
  {V.}~\bibnamefont {Sandoghdar}},\ }\href@noop {} {\bibfield  {journal}
  {\bibinfo  {journal} {Phys. Rev. Lett.}\ }\textbf {\bibinfo {volume} {108}},\
  \bibinfo {pages} {093601} (\bibinfo {year} {2012})}\BibitemShut {NoStop}%
\bibitem [{\citenamefont {Kok}\ and\ \citenamefont {Lovett}(2010)}]{Kok2010}%
  \BibitemOpen
  \bibfield  {author} {\bibinfo {author} {\bibfnamefont {P.}~\bibnamefont
  {Kok}}\ and\ \bibinfo {author} {\bibfnamefont {B.}~\bibnamefont {Lovett}},\
  }\href@noop {} {\emph {\bibinfo {title} {Introduction to Optical Quantum
  Information Processing}}}\ (\bibinfo  {publisher} {Cambridge University
  Press},\ \bibinfo {year} {2010})\BibitemShut {NoStop}%
\bibitem [{\citenamefont {Ralph}\ \emph {et~al.}(2015)\citenamefont {Ralph},
  \citenamefont {S\"ollner}, \citenamefont {Mahmoodian}, \citenamefont
  {White},\ and\ \citenamefont {Lodahl}}]{Ralph:15}%
  \BibitemOpen
  \bibfield  {author} {\bibinfo {author} {\bibfnamefont {T.}~\bibnamefont
  {Ralph}}, \bibinfo {author} {\bibfnamefont {I.}~\bibnamefont {S\"ollner}},
  \bibinfo {author} {\bibfnamefont {S.}~\bibnamefont {Mahmoodian}}, \bibinfo
  {author} {\bibfnamefont {A.}~\bibnamefont {White}}, \ and\ \bibinfo {author}
  {\bibfnamefont {P.}~\bibnamefont {Lodahl}},\ }\href@noop {} {\bibfield
  {journal} {\bibinfo  {journal} {Phys. Rev. Lett.}\ }\textbf {\bibinfo
  {volume} {114}},\ \bibinfo {pages} {173603} (\bibinfo {year}
  {2015})}\BibitemShut {NoStop}%
\bibitem [{\citenamefont {Rotenberg}\ \emph {et~al.}(2017)\citenamefont
  {Rotenberg}, \citenamefont {T\"{u}rschmann}, \citenamefont {Haakh},
  \citenamefont {Martin-Cano}, \citenamefont {G\"{o}tzinger},\ and\
  \citenamefont {Sandoghdar}}]{Rotenberg:17}%
  \BibitemOpen
  \bibfield  {author} {\bibinfo {author} {\bibfnamefont {N.}~\bibnamefont
  {Rotenberg}}, \bibinfo {author} {\bibfnamefont {P.}~\bibnamefont
  {T\"{u}rschmann}}, \bibinfo {author} {\bibfnamefont {H.~R.}\ \bibnamefont
  {Haakh}}, \bibinfo {author} {\bibfnamefont {D.}~\bibnamefont {Martin-Cano}},
  \bibinfo {author} {\bibfnamefont {S.}~\bibnamefont {G\"{o}tzinger}}, \ and\
  \bibinfo {author} {\bibfnamefont {V.}~\bibnamefont {Sandoghdar}},\ }\href
  {\doibase 10.1364/OE.25.005397} {\bibfield  {journal} {\bibinfo  {journal}
  {Opt. Express}\ }\textbf {\bibinfo {volume} {25}},\ \bibinfo {pages} {5397}
  (\bibinfo {year} {2017})}\BibitemShut {NoStop}%
\bibitem [{\citenamefont {T\"urschmann}\ \emph {et~al.}(2017)\citenamefont
  {T\"urschmann}, \citenamefont {Rotenberg}, \citenamefont {Renger},
  \citenamefont {Harder}, \citenamefont {Lohse}, \citenamefont {Utikal},
  \citenamefont {G\"otzinger},\ and\ \citenamefont
  {Sandoghdar}}]{Tuerschmann:17}%
  \BibitemOpen
  \bibfield  {author} {\bibinfo {author} {\bibfnamefont {P.}~\bibnamefont
  {T\"urschmann}}, \bibinfo {author} {\bibfnamefont {N.}~\bibnamefont
  {Rotenberg}}, \bibinfo {author} {\bibfnamefont {J.}~\bibnamefont {Renger}},
  \bibinfo {author} {\bibfnamefont {I.}~\bibnamefont {Harder}}, \bibinfo
  {author} {\bibfnamefont {O.}~\bibnamefont {Lohse}}, \bibinfo {author}
  {\bibfnamefont {T.}~\bibnamefont {Utikal}}, \bibinfo {author} {\bibfnamefont
  {S.}~\bibnamefont {G\"otzinger}}, \ and\ \bibinfo {author} {\bibfnamefont
  {V.}~\bibnamefont {Sandoghdar}},\ }\href@noop {} {\bibfield  {journal}
  {\bibinfo  {journal} {Nano Lett.}\ }\textbf {\bibinfo {volume} {17}},\
  \bibinfo {pages} {4941} (\bibinfo {year} {2017})}\BibitemShut {NoStop}%
\bibitem [{\citenamefont {Diniz}\ \emph {et~al.}(2011)\citenamefont {Diniz},
  \citenamefont {Portolan}, \citenamefont {Ferreira}, \citenamefont {G\'erard},
  \citenamefont {Bertet},\ and\ \citenamefont {Auff\`eves}}]{Diniz:11}%
  \BibitemOpen
  \bibfield  {author} {\bibinfo {author} {\bibfnamefont {I.}~\bibnamefont
  {Diniz}}, \bibinfo {author} {\bibfnamefont {S.}~\bibnamefont {Portolan}},
  \bibinfo {author} {\bibfnamefont {R.}~\bibnamefont {Ferreira}}, \bibinfo
  {author} {\bibfnamefont {J.~M.}\ \bibnamefont {G\'erard}}, \bibinfo {author}
  {\bibfnamefont {P.}~\bibnamefont {Bertet}}, \ and\ \bibinfo {author}
  {\bibfnamefont {A.}~\bibnamefont {Auff\`eves}},\ }\href@noop {} {\bibfield
  {journal} {\bibinfo  {journal} {Phys. Rev. A}\ }\textbf {\bibinfo {volume}
  {84}},\ \bibinfo {pages} {063810} (\bibinfo {year} {2011})}\BibitemShut
  {NoStop}%
\bibitem [{\citenamefont {Haakh}, \citenamefont {Faez},\ and\ \citenamefont
  {Sandoghdar}(2016)}]{Haakh:16}%
  \BibitemOpen
  \bibfield  {author} {\bibinfo {author} {\bibfnamefont {H.~R.}\ \bibnamefont
  {Haakh}}, \bibinfo {author} {\bibfnamefont {S.}~\bibnamefont {Faez}}, \ and\
  \bibinfo {author} {\bibfnamefont {V.}~\bibnamefont {Sandoghdar}},\ }\href
  {\doibase 10.1103/PhysRevA.94.053840} {\bibfield  {journal} {\bibinfo
  {journal} {Phys. Rev. A}\ }\textbf {\bibinfo {volume} {94}},\ \bibinfo
  {pages} {053840} (\bibinfo {year} {2016})}\BibitemShut {NoStop}%
\bibitem [{\citenamefont {Walser}\ \emph {et~al.}(2009)\citenamefont {Walser},
  \citenamefont {Zumofen}, \citenamefont {Renn}, \citenamefont
  {G{\"o}tzinger},\ and\ \citenamefont {Sandoghdar}}]{Walser:09}%
  \BibitemOpen
  \bibfield  {author} {\bibinfo {author} {\bibfnamefont {A.}~\bibnamefont
  {Walser}}, \bibinfo {author} {\bibfnamefont {G.}~\bibnamefont {Zumofen}},
  \bibinfo {author} {\bibfnamefont {A.}~\bibnamefont {Renn}}, \bibinfo {author}
  {\bibfnamefont {S.}~\bibnamefont {G{\"o}tzinger}}, \ and\ \bibinfo {author}
  {\bibfnamefont {V.}~\bibnamefont {Sandoghdar}},\ }\href@noop {} {\bibfield
  {journal} {\bibinfo  {journal} {Mol. Phys.}\ }\textbf {\bibinfo {volume}
  {107}},\ \bibinfo {pages} {1897} (\bibinfo {year} {2009})}\BibitemShut
  {NoStop}%
\bibitem [{\citenamefont {Kozyryev}\ \emph {et~al.}(2017)\citenamefont
  {Kozyryev}, \citenamefont {Baum}, \citenamefont {Matsuda}, \citenamefont
  {Augenbraun}, \citenamefont {Anderegg}, \citenamefont {Sedlack},\ and\
  \citenamefont {Doyle}}]{Kozyryev:17}%
  \BibitemOpen
  \bibfield  {author} {\bibinfo {author} {\bibfnamefont {I.}~\bibnamefont
  {Kozyryev}}, \bibinfo {author} {\bibfnamefont {L.}~\bibnamefont {Baum}},
  \bibinfo {author} {\bibfnamefont {K.}~\bibnamefont {Matsuda}}, \bibinfo
  {author} {\bibfnamefont {B.~L.}\ \bibnamefont {Augenbraun}}, \bibinfo
  {author} {\bibfnamefont {L.}~\bibnamefont {Anderegg}}, \bibinfo {author}
  {\bibfnamefont {A.~P.}\ \bibnamefont {Sedlack}}, \ and\ \bibinfo {author}
  {\bibfnamefont {J.~M.}\ \bibnamefont {Doyle}},\ }\href {\doibase
  10.1103/PhysRevLett.118.173201} {\bibfield  {journal} {\bibinfo  {journal}
  {Phys. Rev. Lett.}\ }\textbf {\bibinfo {volume} {118}},\ \bibinfo {pages}
  {173201} (\bibinfo {year} {2017})}\BibitemShut {NoStop}%
\end{thebibliography}

\end{document}